\documentclass[aps,prd,twocolumn, notitlepage,superscriptaddress,10pt,nofootinbib, 
amsmath,amssymb]{revtex4-1}

\usepackage{color,marvosym}
\usepackage{graphicx} 
\usepackage{comment}

\newcommand{\fett}[1]{\boldsymbol{#1}}

\newcommand{\dd}{{\rm{d}}}

\newcommand{\be}{\begin{equation}}
\newcommand{\ee}{\end{equation}}

\newcommand{\nabq}{\fett{\nabla}_{\fett{q}}}
\newcommand{\nabx}{\fett{\nabla}_{\fett{x}}}
\newcommand{\nab}{\fett{\nabla}}

\newcommand{\stuff}{\mathcal{H}_0^2\Omega_{{\rm m}0}}
\newcommand{\myArxiv}[1]{{\tt \href{http://arxiv.org/abs/#1}{{\color{black}[{\color{black}#1}]}}}}

\definecolor{darkred}{rgb}{0.5,0,0}
\definecolor{darkgreen}{rgb}{0,0.5,0}
\definecolor{darkblue}{rgb}{0,0,0.5}

\usepackage{hyperref}
\hypersetup{colorlinks,
linkcolor=darkblue,
filecolor=darkgreen,
urlcolor=darkred,
citecolor=darkblue }

\definecolor{darkgreen}{rgb}{0,0.5,0}

\begin{document}

\title[Lagrangian theory for cosmic structure formation with vorticity: Newtonian and post-Friedmann approximations]{Lagrangian theory for cosmic structure formation with vorticity: \\ Newtonian and post-Friedmann approximations}
\date{\today} 

\author{Cornelius Rampf}
\email{rampf@thphys.uni-heidelberg.de}
\affiliation{Institut f\"ur Theoretische Physik, Philosophenweg 16, D--69120 Heidelberg, Germany}
\affiliation{Institute of Cosmology and Gravitation, University of Portsmouth, Portsmouth PO1 3FX, United Kingdom}

\author{Eleonora Villa}
\email{evilla@sissa.it}
\affiliation{SISSA, via Bonomea 265, 34136, Trieste, Italy}
\affiliation{INFN, Sezione di Trieste, Via Valerio 2, 34127 Trieste, Italy}
\affiliation{Institute of Cosmology and Gravitation, University of Portsmouth, Portsmouth PO1 3FX, United Kingdom}

\author{Daniele Bertacca}
\email{dbertacca@astro.uni-bonn.de}
\affiliation{Argelander-Institut f\"ur Astronomie, Auf dem H\"ugel 71, D--53121 Bonn, Germany}

\author{Marco Bruni}
\email{marco.bruni@port.ac.uk}
\affiliation{Institute of Cosmology and Gravitation, University of Portsmouth, Portsmouth PO1 3FX, United Kingdom}

\begin{abstract}
We study the nonlinear gravitational dynamics of a universe filled with a pressureless fluid and a cosmological constant $\Lambda$ in the context of Newtonian gravity, 
and in the relativistic post-Friedmann approach proposed in paper I [I.~Milillo \textit{et al.}, Phys.\ Rev.\ D {\bf 92},  023519 (2015).].  The post-Friedmann approximation scheme is based on the $1/c$ expansion of the space-time metric and the energy-momentum tensor, and includes nonlinear Newtonian cosmology. 
Here we establish the nonlinear post-Friedmann framework in the Lagrangian-coordinates approach for structure formation. 
For this we first identify a Lagrangian gauge which is suitable for incorporating nonzero vorticity.
We analyze our results in two limits: at the leading order we recover the fully nonlinear Newtonian cosmological equations 
in the Lagrangian formulation, and we provide a space-time metric consistent from the perspective of general relativity. 
We then linearize our expressions and recover the relativistic results at first order in cosmological perturbation theory.
Therefore, the introduced approximation scheme provides a unified treatment for the two leading-order regimes, from the small scales described by Newtonian gravity  
to the large linear scale, where first-order relativistic cosmological perturbation theory gives a very good description of structure formation.
\end{abstract}

\maketitle

\section{Introduction}\label{sec:intro}

The $\Lambda$CDM  model \cite{Peebles:1984ge, Efstathiou:1990xe} provides today the accepted standard {\it concordance} \cite{Tegmark:2003ud} description of our Universe \cite{Ade:2015xua}.
Its late  dynamics is  dominated by a collisionless cold dark matter (CDM) component and a cosmological \mbox{constant $\Lambda$:}
the latter is responsible for the observed acceleration of the cosmic expansion,  while CDM can collapse and form structures.
Baryons, i.e., matter which interacts gravitationally and electroweakly, and radiation (photons and neutrinos), although responsible for many phenomena we can observe directly, 
are nowadays only a minor constituent of the overall energy budget.

This concordance  model is based on general relativity (GR) as the theory of gravity and on assuming homogeneity and isotropy on very large scales, so that the Universe as a whole is described as a Friedman-Lema\^itre-Robertson-Walker (FLRW) space-time and its overall expansion is  parametrized by the cosmic scale factor $a(t)$, governed by the Friedmann equations.
However,  on small enough scales the Newtonian treatment of structure formation
is usually assumed to be a good approximation. 
Then, assuming  a pressureless fluid (dust) description that is valid at sufficiently early times (before shell crossing), the evolution of the CDM component is given by the Euler--Poisson equations. 

Initial conditions for structure formation, even in the Newtonian description, are set at early times by relativistic and electroweak physics
(governed by the set of coupled Boltzmann--Einstein equations) that also leads to the observation of the cosmic microwave background 
(CMB) anisotropies \cite{Ade:2015xua,Penzias:1965wn,Netterfield:2001yq}. These initial fluctuations are very tiny, and  therefore can be well described as perturbations of the FLRW background space-time, usually divided into three irreducible parts: scalar, vector and tensors {\it modes} (the latter only appearing in the relativistic context and describing gravitational waves). 

Specifically relevant to our analysis is the velocity field of matter, which in general can be split into a scalar part, proportional to the gradient of a velocity potential, and a vector part. In Newtonian theory the vorticity is defined as  the curl of the velocity, which is vanishing if the velocity is exactly of the scalar type. Furthermore, as it follows
from the Kelvin circulation theorem, a pressureless fluid which is initially curl free remains curl free (see, \mbox{e.g., \cite{Buchert:1992ya}).}\footnote{Of course, a fluid which is initially curl free will generate vorticities when it enters into the multistream regime (see, e.g., \cite{Hahn:2014lca}). 
This regime is accompanied with multivalued velocities, and this is also one reason why the single-stream fluid description breaks down.} This result generalizes to  GR \cite{Ellis:1971pg,Kodama:1985bj,Roybook} where, however, vorticity is a four-vector and is defined as the antisymmetric part of the covariant derivative of the four-velocity of matter. In addition, the gravitational field (i.e., the space-time metric) in general  also contains a vector part, leading to the relativistic effect of frame dragging. It follows that, although related, the Newtonian and GR vorticity fields are different, in subtle ways that we aim at elucidating here.

In  cosmological perturbation theory \cite{malik&wands,Kodama:1985bj,Villa:2015ppa} only scalar modes are relevant for structure formation at first order, thus vorticity is usually neglected. In addition, 
studying the linearized fluid equations leads to the observation that, 
in an expanding universe, vorticity decays away as $1/a$ (for a review, see e.g.\ \cite{Bernardeau:2001qr}).
However, a nonvanishing vorticity could potentially have  impact on the (early)  gravitational dynamics. 
Furthermore, there is no reason why the initial vorticity should be exactly vanishing, 
expecially considering that nonlinear CMB physics generates vector perturbations which remain constant at early times (recombination), and are of the order of a few 
percent with respect to  second-order scalar perturbations \cite{Beneke:2010eg,Lu:2008ju,Durrer:2016jzq}. Vorticity can also be present at late times and contributes to generate frame dragging, a  purely relativistic gravitomagnetic (vector-type) effect. This  is  produced at leading order, i.e.\ by a purely Newtonian dynamics, in the post-Friedmann (PF) approximation, as shown in \mbox{paper I} \cite{Milillo:2015cva} and computed 
 in Newtonian $N$-body simulations \cite{Bruni:2013mua,Thomas:2015kua}, and also in the $f(R)$ gravity context \cite{Thomas:2015dfa}; cf.\ also \cite{Adamek:2015eda}.
In order to give a self-consistent and complete description we therefore include vorticity in our analysis.

Vorticity has been previously considered by various authors.
In Refs.\ \cite{Buchert:1992ya,BarrowSaich93,Dominguez:2001sf,Buchert:2005xj} it is discussed in a Newtonian setup  how the vorticity is coupled to the nonlinear density enhancement (see also \cite{Serrin1959}). 
A relativistic treatment of this density-vorticity relation was given in \cite{Ellis:1990gi}, and in \cite{Asada:2000am} even for a relativistic fluid with pressure.
References \cite{Christopherson:2009bt,Christopherson:2010dw} consider vorticity generation in a  fluid with pressure \cite{Ellis:1971pg,Roybook}---a topic we do not investigate in this paper. More relevant for the present study is Ref.\ \cite{Asada:1999ba}, where a relativistic coordinate system/gauge  which is  convenient for investigating vorticity is introduced.

The flow of a fluid can be studied, in GR or in Newtonian physics,  either in the Eulerian or Lagrangian representation. The Eulerian formulation makes use of the coordinate system of a  fixed observer, where the observer is studying how the 
streams of matter are clustering \cite{Villa:2015ppa}. This fixed coordinate system is uniquely defined in the absolute Euclidean space of Newtonian physics but it is completely arbitrary in GR.  In the Lagrangian formulation,
by contrast, the observer makes use of a coordinate system which is attached to the matter elements, i.e., the observer is comoving with the fluid. 

In this paper we investigate the  Lagrangian-coordinates approach to cosmological structure formation, in $\Lambda$CDM and taking vorticity into account, in
(1)  the  Newtonian setting and (2) 
within a specific approximation scheme in GR, namely the post-Friedmann framework \cite{Bruni:2013mua,Milillo:2015cva,Thomas:2014aga,Thomas:2015kua,Thomas:2015dfa}. 
Specifically, in the Newtonian part of this paper we derive nonlinear evolution equations for the dynamical fields in the Lagrangian-coordinates formulation, obtaining some new results. 
One of these is the so-called Cauchy invariants, which are known in the more general literature on fluid 
dynamics (e.g., \cite{YakubovichZenkovich2001,Zheligovsky:2013eca,Podvigina}), 
but in the cosmological case only known in the case of vanishing vorticity (see \cite{Ehlers:1996wg,Rampf:2012up,Zheligovsky:2013eca,Rampf:2015mza,Matsubara:2015ipa}).
Another new Newtonian result we obtain is the generalization of the Bernoulli equation in Lagrangian space including vorticity. These equations, together with
the mass conservation equation and Poisson equation, form a complete set of Newtonian equations which can be solved, e.g., 
by using Lagrangian perturbation theory (e.g., \cite{Buchert:1987xy,Buchert:1989xx,Bouchet:1994xp,Bernardeau:2001qr,Rampf:2012xa,Bernardeau:2013oda}).

The second part of this paper deals with the nonlinear relativistic PF framework \cite{Bruni:2013mua,Milillo:2015cva,Thomas:2014aga,Thomas:2015kua,Thomas:2015dfa}, which in essence is 
  a generalization of the post-Minkowski (weak-field) approximation (see e.g., \cite{Gravity}) to the case of a flat FLRW background space-time, together with the fundamental
assumption  that {\it peculiar} velocities are small.\footnote{This should be contrasted with  the standard post-Newtonian (PN) approximation \cite{Weinberg:1972kfs,Gravity,Takada:1997bk,Futamase1988,Tomita:1987ja,Matarrese:1995sb,Hwang:2005mg}; see paper I \cite{Milillo:2015cva} for a detailed discussion.
See also \cite{Kopp:2013tqa} for a double expansion which seems to be highly related to the post-Friedmann approach.}
As shown in paper I \cite{Milillo:2015cva}, in this  framework one directly recovers,  to the leading order in a $1/c$ expansion in the Poisson (or conformal Newtonian) gauge \cite{Bertschinger:1993xt}, the Newtonian cosmological 
 equations in the Eulerian frame (this justifies  the Poisson gauge as the Eulerian gauge of choice in the relativistic context). 
Furthermore it was also shown in paper I \cite{Milillo:2015cva} that, linearizing the equations and with the use of a resummation scheme, one recovers  first-order relativistic cosmological perturbation theory {\it (CPT)} in the same gauge \cite{Ma:1995ey}.
Thus, the PF approach provides a unified  nonlinear framework to study cosmological structure formation from small scales,
where the Newtonian regime is valid, to large scales, where relativistic {\it CPT} is a good approximation. 

The aim of this paper is to introduce a Lagrangian-coordinates formulation of the PF approach, allowing for vorticity. To the leading order in the $1/c$ expansion, therefore, we obtain the Lagrangian-coordinates formulation of the Newtonian fluid equations;  in addition, we recover first-order relativistic perturbation results when we linearize our equations.
As said above, the post-Friedmann approximation scheme naturally incorporates vorticity and describes frame dragging, features that are directly allowed by the Poisson gauge used in  \cite{Bruni:2013mua,Milillo:2015cva,Thomas:2014aga,Thomas:2015kua,Thomas:2015dfa}.  
The  synchronous-comoving-orthogonal gauge (SCO) is commonly used in the literature when investigating general relativistic Lagrangian fluid dynamics (e.g., \cite{Stewart:1994wq,Matarrese:1995sb,Russ:1995eu,Takada:1997bk,Bruni:2013qta,BruniHidalgoWands,Villa:2014aja,Villa:2014foa}),  but
this gauge can only be used for an irrotational fluid, where the simultaneous
conditions synchronous, comoving and orthogonal hold. These conditions become however
incompatible when nonvanishing vorticity is allowed. 
Here, by applying completely general considerations, we thus motivate our gauge choice which we physically link to  {\it a general Lagrangian frame}, i.e., a coordinate system which allows vorticity, and we call the resulting gauge the Lagrangian gauge. We note that this Lagrangian gauge is constructed in such a way that, in the limit of
vanishing vorticity, it becomes identical with the SCO gauge.
Furthermore, we define the Lagrangian gauge from exact properties on the metric that we deduce from the geodesic equation.

This paper is organized as follows. In Sec.\ \ref{sec:Newton} we report the essentials of the Newtonian approach of structure formation. Specifically in Sec.\ \ref{sec:LagNewton}, we introduce a new Lagrangian-coordinates approach in Newton theory, and there we also report new findings such as the Cauchy invariants and the generalized Bernoulli equation.
Thereafter, we switch to a general relativistic description. Specifically, in Sec.\ \ref{sec:LagrangeGauge} we define first the Lagrangian gauge nonperturbatively,
and in Sec.\ \ref{sec:PFEulerian} we first review the PF approach in the Poisson gauge, which serves as our Eulerian approach in the present paper.
To obtain the corresponding Lagrangian-coordinates approach, our strategy is {\it not} to  solve the field equations in the Lagrangian gauge, but instead we perform a gauge transformation from Poisson gauge to Lagrangian gauge (Sec.\ \ref{ssec:L3}).
We choose to proceed in this way to highlight the following physical interpretation of this  gauge transformation: 
fairly similar to the spatial transformation from Eulerian to Lagrangian coordinates
 in Newton theory---which contains the whole dynamical information of the system, i.e., the Newtonian displacement field, 
the outlined gauge transformation amounts to a four-dimensional coordinate transformation involving, to the leading order in $1/c$, the identical Newtonian displacement field in the spatial component of the gauge transformation, and the Newtonian velocity potential in the temporal part of the gauge transformation.
Finally, we conclude in Sec.\ \ref{sec:concl}.

Notation: when it is necessary for clarity we use the subscript E (Eulerian) for the Poisson gauge with coordinates $x^\mu\dot=(\eta,\fett{x})$, and L (Lagrangian) for the Lagrangian gauge with coordinates $q^\mu\dot=(\tau,\fett{q})$. 
Greek indices refer to space-time coordinates, whereas latin indices refer to spatial coordinates.
For notational simplicity, spatial derivatives with respect to the spatial Lagrangian coordinate $q_i$ are abbreviated with a comma ``$,i$'', and for the spatial Eulerian coordinate $x_i$ we use sometimes a slash ``$|i$''.  $\varepsilon_{ijk}$ is the Levi--Civita symbol.
The subscript 1 is used when a variable is evaluated at first order in cosmological perturbation theory.  
Summation over repeated indices is assumed.
We make use of conformal time and make use of the conformal metric $\gamma_{\mu\nu}= g_{\mu\nu}/a^2$.

\section{Newtonian fluid equations}\label{sec:Newton}

In the following we first review the Newtonian cosmological fluid equations in Eulerian coordinates.  Then, in Sec.\ \ref{sec:LagNewton}, we introduce a novel Lagrangian-coordinates approach, which allows a nonzero vortical component in the fluid velocity.

Before going into the details, let us point out the limitations of the fluid description:
this is valid only in the so-called 
single-stream regime, i.e.\ the description breaks down when fluid trajectories begin to intersect.
In cosmology, this is referred to as shellcrossing, and it is accompanied by the appearance of caustics with extreme densities; 
in general fluid mechanics, this is usually called a blowup. After a possible transition period in a multifluid regime, 
eventually the dynamics of the matter should be given by a much more demanding phase-space description; the evolution of this multistream regime is governed by the 
Vlasov--Poisson equations (see e.g., \cite{Hahn:2014lca,Bernardeau:2013oda}). In this paper we do not investigate  such a phase-space description, which means that our description breaks down when the first shellcrossing occurs.
Yet, our approach allows us to include vorticity in the initial conditions.

\subsection{Newtonian Eulerian approach}\label{sec:EulNewton}

Newtonian physics is based on absolute space, with a Euclidean geometry, and absolute time $t$. In cosmology it is convenient to use the fixed comoving coordinates of the expanding FLRW background; these are the coordinates that in this context are referred to as Eulerian. 
This homogeneous isotropic FLRW universe has matter density 
$\bar \rho$, satisfying a background continuity equation, and its 
 evolution  is described by the  scale factor $a(t)$,  governed by the Friedmann equations.
After subtraction of these background equations,
in Eulerian coordinates 
the Euler equations and the continuity equation for the inhomogeneous cosmic fluid are, respectively,
\begin{align} 
 \partial_\eta \fett{v} + (\fett{v} \cdot \nab ) \fett{v}  &= - {\cal H} \fett{v} + \nab U_{\rm N} \,, \label{EulerEta} \\
 \partial_\eta \delta + \nab \cdot \left[ \left( 1 +\delta \right) \fett{v} \right] &= 0 \,, \label{ContiEta}
\end{align}
where $\fett{v}$ 
is the peculiar velocity, 
$\eta$ the conformal time satisfying $a \dd \eta = \dd t$, and
${\cal H} \equiv (\partial_\eta a)/a$ the conformal Hubble parameter. The system of equations governing the dynamics of the self-gravitation fluid is closed by the Poisson equation for 
$U_{\rm N}$, the cosmological (or peculiar) potential,
\be
  \nab^2 U_{\rm N} = -4\pi G  \bar \rho a^2 \delta \,, \label{Poissonequation}
\ee
where  $\delta = (\rho - \bar \rho)/\bar \rho$ is the matter density contrast.

In this paper we allow the velocity to have a longitudinal and transverse component; 
thus we have \mbox{$\fett{v} = \fett{\nabla} \Phi + \fett{\nabla} \times \fett{A}$,} with $\fett{A}$ being subject to the Coulomb gauge condition $\nab \cdot \fett{A}=0$. 
We define the Newtonian vorticity as 
\be \label{vortdef}
 \fett{\omega} \equiv  \frac 1 2 \fett{\nabla} \times \fett{v} \,,
\ee
or, explicitly, in a Eulerian coordinate system and in index notation
\be \label{vortE}
  \omega^{i}_{\rm E} =  
 \frac{1}{2\sqrt{h}} \varepsilon^{ijk} v_{k|j}^{\rm E} \,,
\ee
where $h=\det[h_{ij}]$ is the determinant of the spatial metric $h_{ij}$. In the following we assume that the Eulerian coordinate system is Cartesian, so  that the metric is $\delta_{ij}$ and  $\det[h_{ij}]=\det[\delta_{ij}]=1$.

The velocity has three degrees of freedom, one scalar and two vector parts, for which we obtain evolution equations in Lagrangian space.

\subsection{Newtonian Lagrangian approach}\label{sec:LagNewton}

Let $\fett{q} \mapsto \fett{x}(\fett{q},\eta)$ be the Lagrangian map from the initial position $\fett{q}$ to the Eulerian position $\fett{x}(\fett{q},\eta)$ at conformal time $\eta$. The map satisfies
\be \label{mapDef}
  \dot {\fett{x}} = \fett{v} \,, \quad 
  \fett{x}(\fett{q},\eta) = \fett{q} + \fett{\cal S}(\fett{q},\eta) \,, \quad
  \fett{x}(\fett{q},\eta_{\rm ini}) = \fett{q} \,,
\ee
where $\fett{\cal S}$ is the Lagrangian \textit{displacement field}.   This has a longitudinal and transverse part in Lagrangian space, even in the case of vanishing vorticity, because of the nonlinearity of the Lagrangian map.  

The first expression in~(\ref{mapDef}) is the Lagrangian representation of the fluid velocity, 
which makes use of the Lagrangian convective time derivative (equivalent with the total time derivative), 
\be
 \dot{\quad} \equiv  \frac{\partial}{\partial \eta} \Big|_{\fett{q}} 
  =  \frac{\partial}{\partial \eta} \Big|_{\fett{x}} + \fett{v} \cdot \nab \,.
\ee
This ``dot derivative'' commutes with the spatial Lagrangian derivatives $\partial / \partial q^i$.

A fundamental object of the Lagrangian formulation is the Jacobian matrix of the coordinates transformation,
\begin{equation}\label{Jac}
\bar {\cal J}^i_{\phantom{j}j} = \frac{\partial x^i}{\partial q^j} = \delta^i_j + {{\cal S}^i}_{\!,j}\,,
\end{equation}
where ${{\cal S}^i}_{\!,j}$ is the deformation tensor. The invariance of the Euclidean line element $\dd\ell^2=\delta_{ij}\dd x^i\dd x^j=h^{\rm L}_{ij}\dd q^i\dd q^j$ gives  the metric in Lagrangian coordinates:
\be
h^{\rm L}_{ij}= \bar {\cal J}^k_{\phantom{j}i} \bar {\cal J}^l_{\phantom{j}j}  \delta_{kl}\,,
\ee
with determinant $\sqrt{h^{\rm L}}=\bar{\cal J}$, where 
here and in the following, $\bar{\cal J} \equiv \det \left[ \bar {\cal J}^i\,\!_j \right]$ 
is the determinant of the Jacobian matrix,
and when it is necessary for clarity we use the subscript ``E'' for Eulerian and  ``L'' for Lagrangian fields/functions.

From the definition of the vorticity, Eq.\,(\ref{vortdef}),
 and the fact that the velocity transforms under a general coordinate transformation as a vector, $v_k^{\rm L}= \bar {\cal J}_{\phantom{k}k}^l v_l^{\rm E}(\fett{x}(\fett{q},\eta))$, 
it is straightforward to obtain an expression for the vorticity in Lagrangian coordinates,
\be \label{vortL}
 \omega^i_{\rm L} = \left( 2 \bar {\cal J} \right)^{-1} \varepsilon^{ijk} v_{k,j}^{\rm L} =
   \left( 2 \bar {\cal J} \right)^{-1} \varepsilon^{ijk} \dot{  {\bar {{\cal J}}}}^{l}_{\phantom{j}j} {\bar {{\cal J}}}_{lk} \,,
\ee
an expression that we use below.

By contrast, we can express the vorticity in Eulerian coordinates, Eq.\,(\ref{vortE}), as\footnote{%
Let us  briefly outline the derivation of Eq.\,(\ref{vortElag}).
Starting from the vorticity in a Cartesian coordinate system, \mbox{$\omega_{\rm E}^i \equiv \varepsilon^{ijk} v_{k|j}^{\rm E}/2$,} we convert the Eulerian derivative into a Lagrangian one by using the inverse of the Jacobian matrix, 
$\bar {\cal J}_{ij}^{-1} = \varepsilon_i^{\phantom{i}lm} \varepsilon_j^{\phantom{j}pq} \bar {\cal J}_{pl} \bar {\cal J}_{qm} /(2 \bar {\cal J})$. 
We then arrive at \mbox{$\omega_{\rm E}^a =  \varepsilon^{ajk} \varepsilon^{lpq} \varepsilon^{\phantom{j}rs}_j \bar {\cal J}_{rp} \bar {\cal J}_{sq}  \dot{ \bar {{\cal J}}}_{kl} /(4 \bar {\cal J})$,} which after contraction of two Levi-Civita symbols yields~(\ref{vortElag}). See also Ref.\,\cite{Rampf:2012xa}.}
\be  \label{vortElag}
\omega^a_{\rm E}(\fett{x}(\fett{q},\eta),\eta) =  \frac{ {\bar {\cal J}}^a_{\phantom{j}i}}{2\bar {\cal J}}  \varepsilon^{ijk}  \dot{  {\bar {{\cal J}}}}^{l}_{\phantom{j}j} {\bar {{\cal J}}}_{lk} \,,
\ee
which, after inspection of Eq.\,(\ref{vortL}) yields the relation \cite{Asada:1999ba}
\be 
   \omega^a_{\rm E} =  {{\bar {{\cal J}}}^a_{\phantom{a}i}} \, \omega^i_{\rm L} \,, 
\ee
as it should.

Let us now consider the dynamics.
A first   equation is the Lagrangian mass conservation, which is obtained by 
 integrating  Eq.\,\eqref{ContiEta}  and making use of the definition of the Lagrangian map \cite{Buchert:1987xy},
\be \label{LagMassCons}
 \delta = 1 / \bar{\cal J}-1 \,.
\ee

We now derive an evolution equation for the pure vector part of the Euler equation in Lagrangian coordinates, the so-called Cauchy invariants.
To this end we adopt, with slight modifications, 
the procedure outlined in \mbox{Ref.\ \cite{Zheligovsky:2013eca}.}
The left-hand side of Eq.\,(\ref{EulerEta}) is the Lagrangian acceleration, and together with the definition of the Lagrangian map we can rewrite this equation as
\be \label{mixEuler}
   \ddot {\fett{x}} = - {\cal H} \dot{ \fett{x}} +  \nab U_{\rm N}^{\rm L} \,,
\ee
where $U_{\rm N}^{\rm L}= U_{\rm N}^{\rm L}(\fett{q},\eta) \equiv U_{\rm N}(\fett{x}(\fett{q},\eta),\eta)$. 
Note that, on the r.h.s.\ of 
Eq.\,(\ref{mixEuler}), we still have a Eulerian spatial gradient, which can be converted to a Lagrangian one.
 Equation~(\ref{mixEuler}) is then, in index notation,
\be \label{mixEuler2}
 \ddot{ x}_i = - {\cal H} \dot{ x}_i +  \bar {\cal J}_{ij}^{-1} \partial^{q_j} U_{\rm N}^{\rm L} \,,
\ee
where  $\bar {\cal J}_{ij}^{-1} = \partial q_j/\partial x_i$ denotes the inverse of the Jacobian matrix.
Introducing the superconformal time $\zeta$, defined by $ a^2 \dd \zeta \equiv a \dd \eta = \dd t$ \cite{Buchert:1989xx,Rampf:2012xa}, the Hubble drag disappears in Eq.\,(\ref{mixEuler2}), which now reads
\be
 \partial_{\zeta}^2 {x}_i =  {\bar {{\cal J}}}_{ij}^{-1} \partial^{q_j}  U_{\rm N}^{\rm L}  a^2 \,.
\ee
Following \cite{Zheligovsky:2013eca}, we rewrite this as 
\be \label{magic1}
 \partial_\zeta \left[ \left( \partial_\zeta x^l \right)  \bar {\cal J}_{li} \right] =
\partial_{q_i} \left( \frac 1 2 \left| \partial_\zeta {\fett{x}} \right|^2 +   U_{\rm N}^{\rm L}  a^2  \right)\,.
\ee
Taking the Lagrangian curl of this equation  we find
\be \label{magic2}
 \partial_\zeta \left[ \varepsilon^{ijk} \left( \partial_\zeta   {\bar {{\cal J}}}_{\phantom{l}j}^{l} \right)  {\bar {{\cal J}}}_{lk} \right] = 0 \,,
\ee
and integrating in superconformal time then gives 
\be \label{CauchySuper}
 \varepsilon^{ijk} \left( \partial_\zeta  {\bar {{\cal J}}}_{\phantom{l}j}^l \right)  {\bar {{\cal J}}}_{lk} = C^i\,,
\ee
where $C^i$ is an integration constant. 
Expressing this equation in terms of conformal time and evaluating it at initial time (ini) according to~(\ref{mapDef}), by virtue of Eq.\,(\ref{vortL}) this integration constant 
is seen to be related to the initial vorticity, i.e., $C^i= 2\, \omega^i_{\rm ini}a_{\rm ini}$. We thus obtain finally the Cauchy invariants
\be \label{CauchyFinal}
\frac 1 2 \, \varepsilon^{ijk}  \dot{  {\bar {{\cal J}}}}_{\phantom{l}j}^l {\bar {{\cal J}}}_{lk} =  \frac{a_{\rm ini}}{a} \omega^i_{\rm ini} \,, 
\ee
which shows that the vorticity decays away with the Hubble expansion. Locally, however, since $a_{\rm ini} \omega^i_{\rm ini}/a  =  \bar {{\cal J}} \, \omega^i_{\rm L}$  [obtained from Eq.\,(\ref{vortL})] implies that the 
vorticity will inevitably grow near caustic formation, where $\bar {\cal J} \simeq 0$.
Again, $\omega^i_{\rm ini}$ is the initial vorticity which could result from
physics before recombination (cf.\ Ref.\ \cite{Beneke:2010eg}), 
and it is \emph{not} the vorticity which is generated at shellcrossing for which we have that $\bar {\cal J} = 0$.

Equations~(\ref{CauchyFinal}) are our final results for the so-called Cauchy invariants for cosmological fluids, which have, to our knowledge, not yet been reported in the literature.\footnote{For the noncosmological version of the Cauchy invariants, see, e.g., Refs.\ \cite{YakubovichZenkovich2001,Zheligovsky:2013eca,Podvigina}, whereas for the cosmological case but with vanishing vorticity, see Refs.\,\cite{Rampf:2012up,Rampf:2015mza}.
For related expressions using differential forms, see Refs.\ \cite{Ehlers:1996wg,Alles:2015vua}.}
The Cauchy invariants are a set of constructive equations which can be used to determine the transverse components of the Lagrangian displacement field. Indeed, imposing an ansatz for the displacement field in powers of the scale factor, i.e., $\fett{\cal S}= \sum_{n=1}^\infty \fett{\cal S}^{(n)}(\fett{q})\, a^n$, a typical ansatz in Lagrangian perturbation theory \cite{Bernardeau:2001qr}, 
we straightforwardly obtain from Eq.\,(\ref{CauchyFinal}) a relation for the $n$th order Taylor coefficient of the transverse part of the displacement in terms of lower-order coefficients,
\be
  \varepsilon^{ijk}  {\cal S}_{k,j}^{(n)} = \delta^n_1 \frac{2 a_{\rm ini}}{\dot a a} \omega^i_{\rm ini} +\!\!\sum_{0 < m <n} \!\!\! \frac{n-2m}{2n}  \varepsilon^{ijk}  {\cal S}_{j,l}^{(m)} \partial^{q_l} {\cal S}_{k}^{(n-m)} .
\ee
Note that the second term on the rhs of this relation is sourced by both vector and scalar components of the displacement. Thus, to subsequently construct the transverse part of the displacement in a recursive way, one requires also a recursion relation for the scalar part of the displacement (see, e.g., Eq.\,(23) in \cite{Rampf:2015mza}). Then, to obtain the displacement field containing both scalar and vector parts, one has to solve at each order a Helmholtz--Hodge problem. See also Ref.\,\cite{Zheligovsky:2013eca} for further details on recursive solution techniques for the Lagrangian displacement field.

By plugging the Cauchy invariants~(\ref{CauchyFinal}) into the vorticity expression~(\ref{vortElag}), we recover the well-known ``Cauchy integral(s)'' (e.g., \cite{Buchert:1992ya,Asada:1999ba})
\be \label{Cauchyintegral}
  \omega^a_{\rm E}(\fett{x}(\fett{q},\eta),\eta) = \frac{a_{\rm ini}}{a} \frac{ \bar {\cal J}_{\phantom{a}i}^a}{ \bar {\cal J}} \, \omega^i_{\rm ini}  \,,
\ee
which, after inspection of~(\ref{LagMassCons}), explicitly states that the vorticity dynamics are coupled to the density \cite{Buchert:1992ya}. 
As evident from the above analysis, the Cauchy invariants~(\ref{CauchyFinal}) and the Cauchy integrals~(\ref{Cauchyintegral}) are intrinsically related to each other; however we stress again that the former are constructive relations to determine the transverse part of the displacement field, whereas the latter give the Eulerian vorticity at arbitrary times in terms of the Lagrangian map \textit{once the displacement field has been determined}.

We now turn back to the Euler equation in order to derive 
the generalization of the Bernoulli equation in Lagrangian space including vorticity.
To obtain the expression for the scalar part of~(\ref{EulerEta}) in Lagrangian coordinates, we again start with its equivalent~(\ref{magic1}), but now take its Lagrangian divergence which reads 
\be \label{Cauchyscalar1}
\frac 1 a \partial_\eta  \left[ a \,\partial^{q_i} \left\{ \dot x^k \bar {\cal J}_{ki} \right\} \right]
= \nabq^2 \left( \frac 1 2 \left| \dot{ \fett{x}} \right|^2 + U_{\rm N}^{\rm L}  \right) \,,
\ee
where $\nabq^2 \equiv \partial_{q_l} \partial^{q_l}$, and we have converted the temporal derivative back to conformal time.

Now, the curly bracket on the lhs of~(\ref{Cauchyscalar1}) is nothing but the Lagrangian gradient of the Lagrangian velocity potential $\Phi^{\rm L}$. 
To see this, we begin with the Helmholtz--Hodge decomposition of the Eulerian 
velocity 
$\fett{v}^{\rm E} \equiv \fett{v}^{\parallel\,\rm E} + \fett{v}^{\perp\,\rm E}$, 
and focus on its longitudinal part which, in index notation, is $v_k^{\parallel\,\rm E} \equiv \partial_{x_k} \Phi^{\rm E}$.
We transform this longitudinal part to (pseudo) Lagrangian space:  $\partial_{x_k} \Phi^{\rm L} = \dot{ x}_k$. All fields and dependences are Lagrangian, but there is one Eulerian spatial gradient left, which, however can be converted into a Lagrangian one by
multiplying the last expression by $\bar {\cal J}_{ki}$. It is then evident that 
\be
  \nabq^2 \,\Phi^{\rm L} =  \partial^{q_i} \left\{ \dot{ x}^k  \bar {\cal J}_{ki} \right\} \,, \label{gradientLagVelPot}
\ee 
which concludes the proof.\footnote{To our knowledge, the expression for the Lagrangian velocity potential, Eq.\,(\ref{gradientLagVelPot}), has not been reported so far in the literature.
The Lagrangian velocity potential, to second order, has been derived in Ref.\,\cite{Rampf:2013dxa}; see their Eq.\,(A17).
It is interesting to note that $\Phi^{\rm L} = \nabq^{-2} \, \partial^{q_i} \left(\dot x^k  \bar {\cal J}_{ki} \right)$ contains (higher-order) contributions of the longitudinal \textit{and transverse part} of the displacement (even if vanishing vorticity is assumed).} 

Taking the inverse Lagrangian Laplacian on Eq.\,(\ref{Cauchyscalar1}), 
the resulting integration constant $c(\eta)$ may be discarded and
we arrive at the generalized Bernoulli equation in Lagrangian space
\be \label{BernoulliL}
  \dot{ \Phi}^{\rm L} + {\cal H} \Phi^{\rm L} - \frac 1 2 \left| \dot{ \fett{x}} \right|^2 = U_{\rm N}^{\rm L} \,.
\ee
We remark again that our derivation of this equation includes vorticity; to our knowledge this is a new result in the cosmological literature (for the version valid for irrotational motion, see Eq.\,(93) in~\cite{Matarrese:1995sb}). In the more general literature on fluid mechanics, a highly related result has been reported in Ref.\ \cite{WuMaZhou2006} (cf.\ their Eq.\,(3.155a) on page 115). 

Equations~(\ref{CauchyFinal}) and (\ref{BernoulliL}) can be used to derive respectively the vector and scalar part of the velocity. Supplementing these equations with the Lagrangian mass conservation and the Poisson equation, we have established a closed set for the Newtonian fluid equations. We note that the Poisson equation~(\ref{Poissonequation}) is still formulated in Eulerian space, whereas the other equations of the closed set are in Lagrangian space. Thus, the closed set of equations provided is in a seemingly mixed form which however could be easily rectified, e.g., by transforming the Poisson equation to Lagrangian space.

\section{Lagrangian frame in GR including vorticity}\label{sec:LagrangeGauge}

For the purposes of this section it is useful to keep in mind the perspective of the 3+1 formalism \cite{Smarr:1977uf}, where the space-time is split in a family of three-dimensional hypersurfaces where the time is constant, the space, plus the time direction, in strict analogy with the Newtonian treatment and with our intuition. 
This geometrical structure, called time slicing or foliation, defines the normal vector field, $n^\mu$, which is  by definition orthogonal to every hypersurface. In addition, the description of the gravitational dynamics introduces the observer's vector field, $t^\mu$, along which the spatial coordinates are constant, meaning that $t^i=0$ by definition. The $0i$ component of the space-time metric, the so-called shift, represents the rate of deviation of the constant-space coordinates field $t^\mu$ from the normal vector field $n^\mu$. In other words, the normal and the observer's vector field coincide only if $g_{0i}= 0$. 
In addition to the geometrical description of the space-time, we have of course another fundamental vector field, namely the four-velocity of the matter, $u^\mu$.
In the Lagrangian approach to the fluid flow, the dynamics is described with respect to a coordinate system attached to the matter elements. The observer is comoving with the fluid and makes use of spatial coordinates such that the fluid is at rest; thus both the spatial components of the observer's vector field and the spatial velocity of the fluid vanish. In the 3+1 formalism we then set $t^i=u^i=0$ for a Lagrangian frame.
Traditionally, in the cosmological literature one often deals with an irrotational flow, and it turns out that the SCO gauge 
is an excellent choice to study such a relativistic Lagrangian fluid flow; see e.g. \cite{Matarrese:1995sb, Bruni:2013qta, Villa:2015ppa}. In this  case, in virtue of the irrotational assumption, the matter four-velocity is a hypersurface-orthogonal vector, and it coincides with the normal field \cite{Ellis:1971pg}. 
In other words, the spatial coordinates of the SCO gauge are constant along the observer's vector field, the normal vector field, and along the world lines of the matter elements; thus $u^i=n^i=t^i=0$. In this case the spatial coordinates are named comoving orthogonal \cite{malik&wands, Kodama:1985bj}.
This choice implies that the shift is vanishing, $g_{0i}=0$. This is possible only in virtue of the irrotational assumption on the matter four-velocity, that is therefore a hypersurface-orthogonal vector and can coincide with the normal field \cite{Ellis:1971pg}. 
Also, in the SCO gauge the time coordinate coincides with the proper time of the fluid and thus $g_{00}=-a^2$ (in conformal time). This is possible only if the fluid is pressureless, i.e., only for dust \cite{Ellis:1971pg}.  This  choice for spatial and temporal coordinates holds only for irrotational dust, as is well known.
In this paper, by contrast, although we still consider dust, we allow a nonzero vorticity in the fluid motion, which makes it impossible to use the SCO gauge. The aim of this section is to construct a gauge that allows for a Lagrangian description in GR in this particular case. 

As we recalled above, a Lagrangian frame is naturally comoving, i.e., the coordinate system attached to the observer is following the fluid and, as a consequence, the coordinate spatial velocity of the fluid vanishes.
However the dust velocity field is not hypersurface orthogonal and we have a nonvanishing shift in the space-time metric. We show that the shift is strictly related to vorticity.
In GR, time is not absolute, but very similar to the spatial Lagrangian coordinate which is a constant label of an individual matter element, there exists also a temporal coordinate which serves as a unique Lagrangian label, i.e., the proper time along the flow lines of matter. For our definition of the Lagrangian frame, we thus make use of this proper time, which we denote $\tau$.
In the following we describe in detail our choice of the coordinate system, and introduce the respective gauge conditions. We call the resulting gauge the \textit{Lagrangian gauge} (for vanishing vorticity, this Lagrangian gauge reduces to the SCO gauge).
Let us then start with the definition of the dimensionless four-velocity, 
\be \label{4veldef}
  u^\mu \equiv \frac{\dd x^\mu}{c \dd \tau} \,,
\ee
where $\tau$ is the proper time along the fluid and the four-velocity is subject to the usual normalization condition $g_{\mu\nu}u^\mu u^\nu = -1$. To make the connection to Newtonian physics as close 
as possible, we define the spatial peculiar velocity to be $v^i \equiv \dd x^i /\dd \eta$, and note that $v^i = \delta^{ij} v_j$. Therefore we have the following relation between the spatial components of the four-velocity and the three-velocity
\begin{equation}
u^i =\frac{\dd x^i}{c \dd \tau}=\frac{\dd x^i}{c \dd \eta}\frac{\dd\eta}{\dd \tau}=\frac{v^i}{ac}u^0\,,
\end{equation}
where $u^0=\dd\eta/\dd\tau$ by definition.
We are now ready to write the conditions defining the Lagrangian gauge. It is evident that our choice for the time coordinate fixes the time component of the four-velocity to be $u^0=1/a$ in conformal time, provided that we consider a pressure-less fluid (which we do). The spatial coordinates are constant along the fluid, thus this implies $v^i=u^i=0$.
Note that by now, we have introduced four conditions on the four-velocity, namely one for the scalar $u^0$ and three for the vector $u^i =  v^i u^0/(ca)$.
These four conditions are not independent, since the components of the four-velocity are constrained by the normalization condition. 
Thus, this gauge is not yet entirely fixed. We return to this point shortly. The four conditions above define rather a class of gauges that can all be called Lagrangian. 
Let us first derive some general relations that hold in any gauge belonging to this class, including the SCO gauge (but only in the irrotational case) and the gauge we are looking for. First of all, note that the conditions we fix on the components of the four-velocity together with the normalization condition imply $g_{00_{\rm L}} = -a^2$.
The expression for the shift can be obtained by exploiting the pressure-less assumption on the matter. Dust moves along geodesics; thus the four-velocity satisfies 
\begin{equation} \label{geodesics}
  u^\sigma \, {u^\mu}_{;\sigma}=0\,,
\end{equation}
where  the semicolon denotes the covariant derivative.
From the expressions of the temporal and spatial components of the geodesic equations in a coordinate system where $v^i=0$ and $u^0=1/a$, it is straightforward to find respectively the following results for the Christoffel symbols
\begin{equation}
 \Gamma^0_{00}= \frac{\cal H}{c} \qquad\qquad \text{and} \qquad\qquad \Gamma^i_{00}=0\,.
\end{equation}
By substituting these in the very definition $\Gamma_{i00} = g_{\mu i} {\Gamma^\mu}_{00}$ which reads in the present case 
\begin{equation}
 \Gamma_{i00}=g_{0i}\Gamma^0_{00}=\frac{1}{2} \left(2 \dot g_{0i} - g_{00,i}\right)\,,
\end{equation} 
(the dot denotes the conformal time derivative in Lagrangian space, and the comma ``$,i$'' denotes a Lagrangian partial derivative with respect to   spatial coordinates $q_i$), 
we obtain a differential equation for the conformal shift $\gamma_{0i}\equiv g_{0i}/a^2$,
\begin{equation}
 \dot \gamma_{0i}+{\cal H} \gamma_{0i}=0 \,.
\end{equation}
The solution of this is decaying as
\begin{equation}
\gamma_{0i}= \frac{C_i}{ca}\,,
\end{equation}
where $C_i$ is a space-dependent constant, and we have added a factor of 1/c 
such that $C_i$ has the dimension of a velocity. 
The shift $g_{0i}=a^2\gamma_{0i}=a C_i/c$ in the space-time metric is responsible for the frame dragging. We come back to the frame dragging in the Lagrangian gauge in Sec.\ \ref{PFLagrange-linear}.

We have just shown that in presence of dust only, and if we use a comoving coordinate system (comoving in space and time),  
the shift in the metric depends on the constant $C_i$. To see the physical meaning of this constant, 
let us consider the relativistic vorticity tensor which is covariantly defined as 
\begin{equation}
\omega_{\mu\nu}= P^\alpha_\mu P^\beta_\nu \nabla_{[\alpha}u_{\beta]}\,,
\end{equation}
where $P_{\mu\nu} = g_{\mu\nu} +u_\mu u_\nu$ is the projection operator in the fluid rest frame, i.e. $P_{\mu\nu}u^\nu=0$. In a comoving coordinate system where $u^i=0$ the covariant components of the relativistic vorticity are purely spatial\footnote{This is not the case for the mixed and contravariant components.} and read 
\begin{equation}
 \omega_{ij} = P^k_iP^n_j u_{[n;k]} = \frac{1}{2}P^k_iP^n_j \left( u_{n ;k} - u_{k ;n} \right) \,,
\end{equation}
where $u_{i}=g_{i\lambda}u^\lambda$.    
Using again the comoving condition $u^i=0$ we find
\begin{align}
\omega_{ij}&=\frac{u^0}{2}P^k_iP^n_j\left(g_{k\lambda}\Gamma^\lambda_{0n}-g_{n\sigma}\Gamma^\sigma_{k 0}\right) \\
&=  u^0 P^k_iP^n_j\left( g_{k0,n}- g_{n0,k}\right) \,.
\end{align}
If, in addition, $u^0=1/a$ and therefore $g_{0i}=a^2\gamma_{0i}=a C_i/c$ (as in our case) we finally have
\begin{equation}\label{vortFD}
\omega_{ij}= P^k_iP^n_j \frac 2 c C_{[n,k]}\,,
\end{equation}
i.e., the time dependence of the vorticity in these coordinates is embodied in the projector tensor.

 Equation~(\ref{vortFD}) shows that the transverse part of the space-dependent constant in the shift represents therefore
a frame-dragging vector potential for relativistic vorticity and cannot be set to 0 in general. 
 Only for the case of irrotational dust, the relativistic vorticity vanishes and it is possible to fix $C_i=g_{0i}/(ac)=0$. In other words, our Lagrangian gauge is comoving but not orthogonal.
Nevertheless, recalling that the conditions $v^i=0$ and $u^0=1/a$ leave us with one more degree of freedom to fix, we choose to set the scalar part of $C_i$ to 0; thus
\be \label{gaugefix}
  C_i \equiv C_i^\perp \,, \qquad \text{with} \quad C_i^{\perp\,,i} =0 \,, 
\ee
where the spatial partial derivative is lowered and raised with the Lagrangian metric $h_{ij}$.  
Equivalently to using the conditions~(\ref{gaugefix}), we could also impose the vanishing of the scalar part of $g_{0i}$, thus leaving the shift to be purely transverse. We call the resulting gauge the Lagrangian gauge.  
As mentioned earlier, in the absence of vorticity, the transverse shift in this gauge then vanishes, and in this limit the Lagrangian gauge corresponds
to the SCO gauge.

Summarizing, we define the Lagrangian gauge by the conditions
\begin{align}
 u_{\rm L }^0 &= \frac 1 a \,, \qquad
 u_{\rm L }^i = 0 \,, \qquad  g_{{0i}_{\rm L}}^{\;\;\;,i} =0 \,.
 \end{align}
These conditions imply, since $u_{0_{\rm L }}=g_{00}u^0_{\rm L}$ and $ u_{i_{\rm L }}=g_{0i}u^0_{\rm L }$, that
\begin{align}
 u_{0_{\rm L }} &= -a \,, \qquad 
 u_{i_{\rm L }} = \frac{C_i^\perp}{c} \,,
\end{align}
where the space-dependent constant $C_i^\perp$ is purely transverse. The components of the metric tensor are given by
\begin{align}
 g_{00_{\rm L }} &= -a^2 \,, \qquad
 g_{0i_{\rm L }} = a \frac{C_i^\perp}{c} \,, \qquad \label{gaugecondshift}
 g_{ij_{\rm L }} = a^2 \gamma_{ij}\,, 
\end{align}
where the spatial metric $g_{{ij}_{\rm  L}}$ contains two scalar, two vector and two tensor degrees of freedom.

The same choice for the space-time coordinates was introduced in Ref.\ \cite{Asada:1999ba} for the study of fluid dynamics in the presence of vorticity in relativistic cosmology. Note however that the analysis of Ref.\ \cite{Asada:1999ba} is restricted to first order in perturbation theory, whereas our analysis is fully nonlinear.
When we expand our results to first order, however, our coordinates and derivatives are exactly the same as in \cite{Asada:1999ba}.

We note that an alternative comoving frame
can be defined using suitable orthonormal coordinates,  Fermi--Walker transported along the world line. This frame is termed Fermi--Walker \cite{ J.L.Synge:1960zz, Misner:1974qy, Felice:2010rpa}. 
The main difference with respect to the Lagrangian frame is that a Fermi--Walker frame is a nonrotating coordinate system, i.e., the 3-space basis is defined by 
three orthogonal axes of a gyroscopes carried by the comoving observer. By construction,  an observer at rest in such a coordinate system cannot measure
the frame dragging, or any other ``Coriolis forces'' which appear e.g.\ in the coordinate transformation from a Eulerian to Lagrangian frame (e.g., \cite{Rampf:2012xa,Rampf:2013dxa}). 
It is also because of that that the Fermi--Walker frame is not a Lagrangian frame.
However, the Fermi--Walker frame could be potentially a valid alternative to the Lagrangian frame of fluid flow, and will be analyzed in a future work.

\section{Post-Friedmann framework: Eulerian-coordinates approach}\label{sec:PFEulerian}

Let us first explain briefly the main difference between the 
PF and PN approximations in cosmology (an extensive discussion about it can be found in paper I \cite{Milillo:2015cva}).
In the PF approach we expand the metric and the energy-momentum tensor in powers of $1/c$ (as in PN approach),  but we keep the matter density and  peculiar velocity  as  exact fundamental variables, assuming however that the latter is small with respect to $c$ (in a PN expansion, the whole physical velocity is assumed to be small, thus a PN approach in cosmology is only valid inside the Hubble horizon).  
The PN expansion is based on an iterative approach; in the PF framework we define a set of resummed PF variables which satisfy consistently both nonlinear evolution and constraint equations.
In section \ref{sec:PFeulerNewton} we see that considering only  the leading order in $1/c$, we recover the Eulerian-coordinates formulation of the Newtonian fluid equations.

Then, in Sec.\ \ref{sec:EulerLinear} we report the linearized evolution equations and derive their solutions. Explicitly, the latter has not been given in paper I \cite{Milillo:2015cva}.
Before proceeding with the 
analysis, let us briefly comment that in general relativity there is no unique coordinate system which could be called Eulerian; 
apart from the Poisson gauge other possible 
gauge choices are for example the harmonic gauge \cite{Szekeres:2000ki}, and, in the context of standard perturbation theory, the
 total matter gauge \cite{Villa:2015ppa}, or the $N$-body gauge \cite{Fidler:2015npa}. However, one advantage of the Poisson gauge is that it remains as close as possible to diagonal and spatially conformally flat, apart from  subdominant  parts, i.e. a spatial transverse-traceless tensor  and a transverse vector.

Let us report some general definitions which are useful in this paper.
The four-dimensional line element (Einstein summation implied)
\be
  \dd s^2 = g_{\mu\nu} \dd x^\mu \dd x^\nu
\ee
with metric signature $(-,+,+,+)$ has, in the Poisson gauge, the following 
metric components in the $1/c$ expansion
\begin{subequations}
\begin{align}
 g_{00} &= - a^2 \left[ 1 -\frac{2U_{\rm N}}{c^2} + \frac{1}{c^4} \left( 2 U_{\rm N}^2 - 4 U_{\rm P} \right) \right]  \,, \label{g00P} \\
 g_{0i} &= - \frac{a^2}{c^3} B_i^{\rm N} - \frac{a^2}{c^5} B_i^{\rm P}   \,, \\
 g_{ij} &= a^2 \left[ \left( 1 + \frac{2 V_{\rm N}}{c^2} + \frac{1}{c^4} \left( 2 V_{\rm N}^2 + 4 V_{\rm P} \right) \delta_{ij} + \frac{1}{c^4} h_{ij}  \right)  \right] \,, \label{gijP}
\end{align}
\end{subequations}
up to $O\left( 1/c^6 \right)$. We set the metric coefficients to be dimensionless. In the Poisson gauge, the vectors in the $g_{0i}$ component are transverse with respect to  to flat space-time, i.e., $ \delta^{ij} B_{i|j}^{\rm N} = 0$, where a slash ``$|j$'' denotes an Eulerian partial derivative with respect to  the spatial component $x_j$,
 and $h_{ij}$ is a transverse and trace-free tensor (${h^i}_i =0 = {h_{ij}}^{|i}$).
In contrast to paper I \cite{Milillo:2015cva}, we make use of the conformal time $\eta$ defined by $a \dd \eta = \dd t$, 
and our time coordinate is $x^0 =c\eta$ (note the factor of $c$, which is an essential aspect of the PF approach). 
The components of the four-velocity are obtained from the metric coefficients~(\ref{g00P})--(\ref{gijP}) and the normalization condition $g_{\mu\nu}u^\mu u^\nu = -1$. We have \cite{Milillo:2015cva}
\begin{widetext}
\begin{subequations}
\begin{align}
 u^0 &= \frac 1 a \left[ 1+\frac{1}{c^2}\left( U_{\rm N} + \frac 1 2 v^2 \right) + \frac{1}{c^4} \left( \frac 1 2 U_{\rm N}^2 + 2U_{\rm P} + v^2 V_{\rm N} + \frac 3 2 v^2 U_{\rm N} + \frac 3 8 v^4 - B^{\rm N}_i v^i  \right) \right] \,,\\
 u^i &=  \frac{v^i}{c \,a} u^0 \,, \\
 u_0 &= a \left[ -1 + \frac{1}{c^2} \left( U_{\rm N} - \frac 1 2 v^2 \right) + \frac{1}{c^4} \left( 2U_{\rm P} -\frac 1 2 U_{\rm N}^2 - \frac 1 2 v^2 U_{\rm N} - v^2 V_{\rm N} - \frac 3 8 v^4 \right) \right] \,, \\
  u_i &=  \frac{a v_i}{c} + \frac{a}{c^3} \left[ -B^{\rm N}_i + v_i U_{\rm N} + 2v_i V_{\rm N} + \frac 1 2 v_i v^2\right]\,,
\end{align}
\end{subequations}
\end{widetext}
where $v^i$ is the spatial peculiar velocity defined by \mbox{$v^i \equiv \dd x^i /\dd \eta$} and $v^2=\delta_{ij}v^iv^j$.
These expressions are the starting point for the calculations in section \ref{ssec:L3}, where we perform the transformation to the Lagrangian coordinates.

\subsection{The Newtonian regime}\label{sec:PFeulerNewton}

Following the notation of paper I \cite{Milillo:2015cva}, 0PF and 1PF orders respectively refer to terms proportional to $1/c^{2}$ and $1/c^{4}$. In particular, the 0PF terms are Newtonian, the 1PF terms contain GR corrections. Here we will consider only the 0PF limit; see paper I \cite{Milillo:2015cva} for the results up to 1PF in the Poisson gauge.

It is easy to see that, retaining the leading-order terms in the $1/c$ expansion from the hydrodynamic equations [Eqs.\,(5.4) and (5.6) in paper I \cite{Milillo:2015cva}], we obtain  the Newtonian continuity and Euler equation, i.e., our Eqs.\,(\ref{EulerEta}) and~(\ref{ContiEta}). Furthermore, from the Einstein 
equations we have \cite{Milillo:2015cva}
\begin{widetext}
\begin{subequations}
\begin{eqnarray}
G_{\;\;0}^0+\Lambda=\frac{8\pi G}{c^4}T^0_{\;\;0} \quad &\rightarrow& \quad \;\frac{1}{c^2}\frac{1}{a^2}\Delta V_{\rm N}=-\frac{4\pi G}{c^2}\bar\rho\delta\;,\\
G^0_{\;\;i}=\frac{8\pi G}{c^4}T^0_{\;\;i}\quad &\rightarrow&\quad \;\frac{1}{c^3}\left[-\frac{1}{2a^2}\Delta B^{\rm N}_{i}+2\mathcal{H} U_{{\rm N}|i}+2\dot V_{{\rm N}|i}\right]=\frac{8\pi G}{c^3}\bar\rho(1+\delta) v_i\;,\\
{\rm trace\; of\;\; } G^i_{\;\;j}+\Lambda \delta^i_{\;\;j}=\frac{8\pi G}{c^4}T^i_{\;\;j} \quad &\rightarrow& \quad \;\frac{1}{c^2}\frac{2}{a^2}\Delta (V_{\rm N}-U_{\rm N})=0\;, \label{traceLinear}\\
{\rm traceless\; part\; of}\;\; G^i_{\;\;j}+\Lambda\delta^i_{\;\;j}=\frac{8\pi G}{c^4}T^i_{\;\;j} \quad &\rightarrow& \quad \;\frac{1}{c^2}\frac{1}{a^2}\bigl[(V_{\rm N}-U_{\rm N})_{|i}^{\;\;|j}-\frac{1}{3}\Delta (V_{\rm N}-U_{\rm N})\,\delta_{i}^{j}\bigr]=0\;,
\end{eqnarray}\\
\end{subequations}
\end{widetext}
where we remind the reader, that a slash ``$|i$'' denotes a Eulerian derivative with respect to  spatial component $x_i$, and $\Delta \equiv \nabx^2$ denotes the Eulerian Laplacian. 
From~(\ref{traceLinear}) it is evident that in this regime $U_{\rm N}=V_{\rm N}$. In other words, Einstein equations reduce  to the standard equations of Newtonian cosmology.
The metric tensor generated  from a self-consistent expansion of the full set of Einstein equations at leading order is the cosmological version of the weak-field metric, with the FLRW metric replacing Minkowski as background. Finally, let us also note that in the Newtonian regime the frame-dragging term, which in the Poisson gauge is represented by $B^{\rm N}_i$, cannot be set to 0.

\subsection{The linear limit}\label{sec:EulerLinear}

The linearization of the PF equations has been investigated in paper I \cite{Milillo:2015cva}. 
The authors define a set of appropriately resummed variables for the scalar sector, which are,  valid to first order in perturbation theory
\begin{align}
\label{ph1i}
 \phi_1 &\equiv - \left(U_{\rm N}+\frac{2}{c^2}U_{\rm P} \right) \,,\\
\label{psi1}
 \psi_1 &\equiv - \left( V_{\rm N}+\frac{2}{c^2}V_{\rm P} \right) \,,
\end{align}
and for the vector sector, given by
\be
\label{vect1}
\omega_{1i} \equiv B^{\rm N}_i+\frac{1}{c^2}B^{\rm P}_i \,.
\ee
Note that in the linear limit, Eulerian temporal and spatial derivatives coincide with the Lagrangian ones.
In paper I \cite{Milillo:2015cva} it is shown that the resummed variables defined above satisfy the same equations as the fully GR equations at first order.
We report in the following the first-order relativistic solutions, which are missing in paper I \cite{Milillo:2015cva}. The line element at first order in the Poisson gauge is given by
\begin{align}
\dd s^2 &= a^2 \Bigg\{-\left( 1+2 \frac{\phi_1}{c^2} \right)c^2\dd \eta^2 - 2\frac{\omega_{1i}}{c^3}c\dd \eta \dd x^i \nonumber \\
   &\quad\hspace{2cm}+\left(1-2\frac{\psi_{1}}{c^2}\right) \delta_{ij}\, \dd x^i \dd x^j \Bigg\} \,,
\end{align}
where the shift is purely transverse, i.e., $\delta^{ij}\omega_{1\,j}=0$. The spatial components of the four-velocity are decomposed in scalar and vector parts, 
\begin{equation}
u^i _{1}= \delta^{ij}v_{1\,,j}+ v^i_{1\perp}\,,
\end{equation}
where $v^i_{1\perp,i}=0$.

The solution for the scalars is well known. We just report here the results of Ref.~\cite{Villa:2015ppa}, to which we refer for the details. The scalars in the metric tensor are found to be equal and given by
\begin{equation}
\psi_{1}=\phi_{1} \equiv g\,\varphi_0(\fett{x})\,,
\end{equation}
where $\varphi_0$ is peculiar gravitational potential linearly extrapolated to the present time $\eta_0$, and the time-dependent function $g :={\cal D}/a$ is the growth-suppression factor, with ${\cal D}$ being the growing mode of the linear density contrast $\delta_{1}(\eta, \fett{x})={\cal D}(\eta)\delta_{1}(\fett{x})$. 

The solution for the scalar part of the spatial four-velocity is
\begin{equation}
v_{1}  =- \frac{2}{3 {\cal H}_0^2 \Omega_{{\rm m}_0}} \dot {\cal D} \varphi_{0} \,, \label{v1P} 
\end{equation}
where $\mathcal{H}_0$ and $\Omega_{{\rm m}0}$ are the Hubble parameter and the matter density parameter $\Omega_{\rm m} \equiv 8 \pi G a^2 \bar \rho/ (3{\mathcal H}^2)$ evaluated at present time.

We now derive the solutions for the first-order shift and for the vector component of the spatial velocity, representing the frame dragging and the vorticity in the matter flow, respectively.
We begin with the $0i$ and trace-free $ij$ component of the field equations in the Poisson gauge, which are for the vector part respectively
\begin{align}
  \frac{1}{c^3} \Delta \omega_{1i} &=  - \frac{1}{c^3}  \frac{6 \stuff}{a}\left( v_{1i_\perp} - \frac{1}{c^2} \omega_{1i} \right) \,, \label{draggingPoiss} \\
 0&= 2 {\cal H} \frac{\omega_{1(i,j)}}{c^4} + \frac{\dot \omega_{1(i,j)}}{c^4} \,. \label{shiftPoissLinear}
\end{align}
From the momentum conservation $T^\mu_{i;\mu}=0$ we obtain an evolution equation for the vector components of the spatial velocity and the shift
\be
  \frac{ \dot \omega_{1i} + {\cal H} \omega_{1i}}{c^4} =   \frac{\dot v_{1i_{\perp}} + {\cal H}  v_{1i_{\perp}}}{c^2} \,,
\ee
whose solution is 
\be
  v_{1i_{\perp}} = \frac{\omega_{1i}}{c^2} + \frac{C_{1i}^\perp}{a} \,,
\ee
 where $C_{1i}^{\perp}$ is a transverse and space-dependent constant. Substituting this result in Eq.\,(\ref{draggingPoiss}) it becomes
\be
   \frac{1}{c^3} \Delta \omega_{1i} =  - \frac{1}{c^3} \frac{6 \stuff}{a^2} C_{1i}^\perp \,. 
\ee
Finally we obtain the following explicit solutions for the vector part of the shift and of the velocity, respectively
\begin{align}
  \omega_{1i} &= - \frac{6 \stuff}{a^2} \Delta^{-1} C_{1i}^\perp (\fett{x}) \,,\label{omega1P} \\
  v_{i1_\perp} &= - \frac{6 \stuff}{a^2 c^2} \Delta^{-1} C_{1i}^\perp(\fett{x}) + \frac{C_{1i}^\perp}{a}  \label{viperp}\,,
\end{align}
where $\Delta^{-1}$ denotes the inverse of the Eulerian Laplacian. 
So far we have not used Eq.\,(\ref{shiftPoissLinear}), but it is easily verified that
solution~(\ref{omega1P}) is in accordance with~(\ref{shiftPoissLinear}).
Let us remark that, to our knowledge, the above results for the frame dragging and the vorticity at first order in perturbation theory are new.

\section{Post-Friedmann framework: transformation to the Lagrangian gauge}\label{ssec:L3}

One aim of this paper is to obtain the PF approximation in the Lagrangian approach. 
To do so, we start from the results for the metric and matter variables in the Poisson gauge obtained in paper I \cite{Milillo:2015cva} (reviewed in section \ref{sec:PFEulerian}), and perform a gauge transformation in terms of a $1/c$ expansion.
The transformation from the Poisson gauge, with coordinates $x^\mu \dot= (\eta,\fett{x})$, to the Lagrangian gauge with coordinates $q^\mu \dot= (\tau,\fett{q})$ is
\begin{align} \label{gaugetrafo}
 \begin{split}
  \eta (\tau,\fett{q}) &= \tau + \frac{1}{c^2} \xi^0(\tau,\fett{q}) + \frac{1}{c^4} \chi^0(\tau,\fett{q})+ O\left(\frac{1}{c^6}\right) \,, \\
   x^i(\tau,\fett{q}) &= q^i + {\cal S}^i(\tau,\fett{q}) +\frac{1}{c^2} \Sigma^i(\tau,\fett{q}) + O \left(\frac{1}{c^4} \right) \,.
 \end{split}
\end{align}

The form of the coordinate transformation is not arbitrary, but it can be easily verified that only the presented transformation does not lead to any inconsistencies.
In particular, we fix the correct powers in $1/c$ by considering the transformation rule for the metric tensor (cf.\ Ref.\,\cite{Gravity}),
\begin{equation} \label{metricTgeneral}
g_{\mu\nu_{\rm L}}(q^\alpha)=\frac{\partial x^\sigma}{\partial q^\mu}\frac{\partial x^\lambda}{\partial q^\nu}g_{\sigma\lambda_{\rm E}}(x^\alpha(q^\rho))\,,
\end{equation}  
where we recall that the subscripts ``L'' and ``E'' indicate respectively Lagrangian gauge and Poisson gauge variables.

In the spirit of Newtonian physics, we define the Eulerian spatial three-velocity in terms of the Lagrangian time derivative of the Lagrangian 
map $\fett{q} \mapsto \fett{x}(\fett{q},\eta)$ by 
\begin{equation} \label{vpoisLrep}
  v_{i_{\rm E}}(x^\mu(q^\nu)) = \frac 1 c \dot{ {\cal S}}_i +\frac{1}{c^3} \dot{ \Sigma}_i + O(1/c^5) \,,
\end{equation}
where $\dot{\quad} \equiv  \partial/\partial \eta |_{\fett{q}}$. We recall that ${\bar{\cal J}^i}\,\!_j$  is the Jacobian matrix element of the spatial Newtonian coordinate transformation $x^i(\tau,\fett{q}) = q^i + {\cal S}^i(\tau,\fett{q})$, namely $\bar{\cal J}^i_{\phantom{j}j} \equiv\delta^i_j+{\cal S}^i_{\phantom{i},j}$, thus as introduced in Sec.\ \ref{sec:Newton}.

Some comments are in order before continuing with the calculations.
The coordinate transformation~\eqref{gaugetrafo} mixes the $1/c$ powers of the variables in the original gauge, because of the additional $1/c$ factor which comes together with the temporal derivative. Therefore we decide to start from the standard PN expansion of the metric and fluid variables in the Poisson gauge and we perform the transformation iteratively in powers of $1/c$. We finally resum the results in the Lagrangian gauge to obtain the 0PF order and the linear PT. 

To obtain the equations for the time gauge generator and for the shift in the Lagrangian gauge, we use the transformation of the components of the four-velocity, which are
\begin{subequations}
\be
 u_{0_{\rm L }}(q^\alpha)  = \frac{\partial x^0}{\partial q^0}u _{0_{\rm E }}(x^\alpha(q^\rho)) + \frac{\partial x^i}{\partial q^0} u_{i_{\rm E }}(x^\alpha(q^\rho)) \,, \label{trasfou0} 
\ee
\be
  u_{i_{\rm L }}(q^\alpha) = \frac{\partial x^0}{\partial q^i} u_{0_{\rm E }}(x^\alpha(q^\rho)) + \frac{\partial x^l}{\partial q^i} u_{l_{\rm E }}(x^\alpha(q^\rho)) \,. \label{trasfoui}
\ee
\end{subequations}

We now expand in $1/c$ powers the Jacobian matrix and the arguments of all the Eulerian variables in the last two equations according to the coordinate transformation~\eqref{gaugetrafo}.
After straightforward calculations we find for Eq.\,\eqref{trasfou0} up to $O(1/c^4)$
\begin{widetext}
\begin{subequations}
\begin{align}
  0 &= \frac{1}{c^2} \left(  U_{\rm N} - {\cal H} \xi^0 - \dot{ \xi}^0 - \frac 1 2 v^2 + v_{i} \, \dot{ {\cal S}}^i  \right) + \frac{1}{c^4} \Bigg[  2U_{\rm P}  - {\cal H} \chi^0  - \dot{ \chi}^0 - \frac 1 2 ( \dot {\cal H} + {\cal H}^2 ) {\xi^0}^2  + ( \dot \xi^0 + {\cal H} \xi^0 ) \left( U_{\rm N} - \frac{ v^2}{2} \right)   
     \nonumber \\ &  + \dot \Sigma^i v_i-\frac 1 2 U_{\rm N}^2-\frac 1 2 v^2U_{\rm N}-v^2V_{\rm N}-\frac 3 8 v^4  
     - \dot \xi^0 {\cal H} \xi^0   + \dot {\cal S}^i \left(  v_i {\cal H} \xi^0  -B^{\rm N}_i+v_iU_{\rm N}+2v_iV_{\rm N}+\frac 1 2 v_iv^2 \right) 
      \Bigg]\,, \label{u0trafo}
\end{align}
and for Eq.\,\eqref{trasfoui} we obtain up to $O(1/c^5)$
\begin{align}
   \frac{C_{i}^\perp}{c a} &= 
   \frac{1}{c} \left( \bar{\cal J}^l{}_{i} v_{l} - \partial_{q_i} \xi^0 \right)  
   + \frac{1}{c^3} \Bigg[ \partial_{q_i} \xi^0 \left( U_{\rm N} - \frac 1 2 v^2 - {\cal H} \xi^0 \right) 
  + \bar{\cal J}^l{}_i \left( v_{l} {\cal H} \xi^0 - B_l^{\rm N} + 3v_{l} U_{\rm N} + \frac 1 2 v_{l} v^2  \right)
    +v_{l} \partial_{q_i} \Sigma^l - \partial_{q_i} \chi^0
      \Bigg] . \label{uitrafo}
\end{align}
\end{subequations}
In the last two equations the time dependence is on the absolute Newtonian time $\eta = \tau$ and the Eulerian variables $U_{\rm N}$, $U_{\rm P}$, and $B_i^{\rm N}$ depend on the Newtonian spatial coordinates $x^i=q^i + {\cal S}^i(\tau,\fett{q})$.

Let us now obtain the expression for the metric tensor in the Lagrangian gauge which is found from the transformation rule~(\ref{metricTgeneral}).
The transformations of the $00$ and $0i$ components, together with the gauge conditions $g_{00_{\rm L }} = -a^2$ and 
 $g_{0i_{\rm L }} = a C_{i}^\perp/c$, lead to the identical equation that we obtain when expanding $u_{0_{\rm L }}$ and $u_{i_{\rm L }}$ [Eqs.\,\eqref{u0trafo} and \eqref{uitrafo}].
Now we proceed with the calculation of the spatial metric in the Lagrangian gauge. 
The transformation rule for the spatial metric reads
\begin{align}
g_{ij_{\rm L}} &= \frac{\partial x^0}{\partial q^i}\frac{\partial x^0}{\partial q^j}g_{00_{\rm E}} +\left(\frac{\partial x^0}{\partial q^i}\frac{\partial x^k}{\partial q^j}+\frac{\partial x^k}{\partial q^i}\frac{\partial x^0}{\partial q^j}\right)g_{0k_{\rm E}}+\frac{\partial x^k}{\partial q^i}\frac{\partial x^n}{\partial q^j}g_{kn_{\rm E}}\,.
\end{align}
The expansion in powers of $1/c$ gives, up to $O(1/c^4)$, 
\begin{align}\label{gijtrafo}
g_{ij_{\rm L}} &= a^2\bar{\cal J}^k\,\!_i \bar{\cal J}^n\,\!_j \delta_{kn} +\frac{a^2}{c^2}\left[ 2\Sigma^k_{,(i} \bar{\cal J}_{kj)} -\xi^0_{,i} \xi^0_{,j} +\left(2V_{\rm N}+2\mathcal{H}\xi^0\right)\bar{\cal J}^k\,\!\!_i\bar{\cal J}^n\,\!\!_j \delta_{kn} \right]   \nonumber \\ 
& +\frac{a^2}{c^4}\Bigg\{  \xi^0_{,i} \xi^0_{,j} \left(2U_{\rm N}-2\mathcal{H}\xi^0\right)- 2 \xi^0_{,i}\bar{\cal J}^k\,\!_j B^{\rm N}_k -  \xi^0_{,(i}\chi^0_{,j)} 
 +4  \Sigma^k_{,(i} \bar{\cal J}^n\,\!_{j)} \left( V_{\rm N} + \mathcal{H}\xi^0\right)\delta_{kn}
\nonumber \\
&  + \bar{\cal J}^k\,\!_i \bar{\cal J}^n\,\!_j\left[\left(2V_{\rm N}^2+4V_{\rm P}+4\mathcal{H}\xi^0 V_{\rm N}+ 2\xi^0 \dot{ V}_{\rm N}+2 \frac{\partial V_{\rm N}}{\partial x^i}\Bigg|_{x^i=q^i+{\cal S}^i} \Sigma^i+\mathcal{H}\xi^{0^2}+\mathcal{H}\chi^0\right)\delta_{kn}+h_{kn}\right]  \Bigg\} \,,
\end{align}
\end{widetext}
where a comma ``$,i$'' denotes a partial spatial derivative with respect to  Lagrangian coordinate $q_i$, as usual. In Eq.\,(\ref{gijtrafo}) the time dependence is on the absolute Newtonian time $\eta = \tau$, and the spatial dependence of the Eulerian functions is on the Newtonian coordinates $x^i=q^i + {\cal S}^i(\tau,\fett{q})$. 

After having obtained the equations for the transformation to the Lagrangian gauge,
we now describe the procedure to solve them for the gauge generator and the space-time metric. First of all, we consider the very definition of the spatial velocity in terms of the map  $\fett{q} \mapsto \fett{x}(\fett{q},\eta)$, which we gave in Eq.\,(\ref{vpoisLrep}) and repeat here for convenience:  $v_{i_{\rm E}}(x^\mu(q^\nu)) = \dot{ {\cal S}}_i/c +\dot{ \Sigma}_i/c^3 + O(1/c^5)$. Once the velocity in the Eulerian coordinates is known, e.g., by the use of cosmological perturbation theory, the displacement field is easily obtained by time integration. Then, Eq.\,\eqref{u0trafo} and \eqref{uitrafo} are four coupled equations which form a close set for the time gauge generator and for the shift in the Lagrangian gauge, $g_{0i_{\rm L }} = a C_{i}^\perp/c$. Finally, the substitution of these results in Eq.\,\eqref{gijtrafo} yields the spatial metric $g_{ij_{\rm L}}$, thus concluding the strategy to derive the Lagrangian metric.

\subsection{The Newtonian regime}

The Newtonian limit is given by the lowest order in the $1/c$ expansion. Equations~(\ref{u0trafo}) and~(\ref{uitrafo}) give respectively
\begin{subequations}
\begin{align}
 \dot{ \xi}^0 + {\cal H} \xi^0 - \frac 1 2 v^2_{\rm E} &=U_{\rm N}   \,,\label{BernNewtlimit} \\ 
 \frac{C_{i}^\perp}{a}&= \bar{\cal J}^l{}_{i} v_{l_{\rm E}} - \partial_{q_i} \xi^0 \,, \label{VelocityNewt}
\end{align} 
\end{subequations}
where in the first equation, from the lowest order of Eq.\,(\ref{vpoisLrep}),  we have used the fact that $v_{i_{\rm E}}(x^\mu(q^\nu))  = \frac 1 c \partial_\eta {\cal S}_i + O(1/c^3)$.
Equation~(\ref{BernNewtlimit}) is nothing but the Newtonian Bernoulli equation [Eq.\,(\ref{BernoulliL})], 
\be 
  \dot{ \Phi}^{\rm L} + {\cal H} \Phi^{\rm L} - \frac 1 2 \left| \dot{ \fett{x}} \right|^2 = U_{\rm N}^{\rm L} \,,
\ee
provided that we identify the time-gauge generator $\xi^0$ with the Lagrangian velocity potential $\Phi^{\rm L}$. 
Indeed, taking the Lagrangian divergence on~(\ref{VelocityNewt}), we recover~(\ref{gradientLagVelPot}) which is the Lagrangian gradient of the Newtonian velocity potential in Lagrangian space.
Finally, by taking the Lagrangian curl of Eq.\,(\ref{VelocityNewt}),
and noting that $ v_{\rm E}^{l,k} \equiv \dot{  {\bar {{\cal J}}}}^{lk}$,
 we recover the Cauchy invariants [cf.\ Eqs.\,(\ref{CauchyFinal})], 
\be \label{newtvort}
 \frac 1 2 \varepsilon^{ijk} \dot{  {\bar {{\cal J}}}}_{\phantom{j}j}^{l}  {\bar {{\cal J}}}_{lk}   = \frac{a_{\rm ini}}{a} \omega_{\rm ini}^i \,, \,\, \qquad (i=1,2,3)
\ee
provided that we identify $\varepsilon^{ijk}C_{k,j}^\perp = 2\,a_{\rm ini} \omega_{\rm ini}^i$. We thus conclude that we have recovered the Lagrangian-coordinates approach of Newtonian dynamics.

\subsection{The linear limit} \label{PFLagrange-linear}
Linearizing our expressions~(\ref{u0trafo}) and~(\ref{uitrafo}) we obtain respectively
\begin{subequations}
\begin{align}
  &0 = \frac{1}{c^2} \left( U_{1\rm N} - {\cal H} \xi^0_1 - \dot{ \xi}^0_1  \right)
     + \frac{1}{c^4} \left[ 2 U_{1 \rm P} - {\cal H} \chi^0_1 - \dot{ \chi}^0_1  \right] \,,      \label{u0trafo1} \\
 &  \frac{C_{1i}^\perp}{c\,a} =
    \frac 1 c \left( \dot{\cal S}_{1i}  -  \xi^0_{1,i}  \right)
    + \frac 1 {c^3} \left[ \dot \Sigma_{1i} - B_{1i}^{\rm N} -  \chi^0_{1,i}  \right] \,,  \label{uitrafo1}
\end{align}
\end{subequations}
where we have used  again Eq.\,(\ref{vpoisLrep}), i.e., $v_{i_{\rm E}}(x^\mu(q^\nu)) = \dot{{\cal S}}_i /c +\dot \Sigma_i/{c^3} + O(1/c^5)$.
The linearization of  Eq.\,\eqref{gijtrafo} leads to the first-order spatial metric
\begin{align} \label{gijtrafo1}
g_{1ij_{\rm L}}&=a^2\Bigg\{\left[1+2\left(\frac{V_{1\rm N}}{c^2}+\frac{2V_{1\rm P}}{c^4}+\mathcal{H}\frac{\xi^0_1}{c^2}+\mathcal{H}\frac{\chi^0_1}{c^4}\right)\right] \delta_{ij} 
 \nonumber \\  &+\left({\cal S}_{1i}+\frac{1}{c^2}\Sigma_{1i}\right)_{,j} + \left({\cal S}_{1j}+\frac{1}{c^2}\Sigma_{1j}\right)_{,i} \Bigg\} \,.
\end{align}
Following paper I \cite{Milillo:2015cva}, we introduce the resummed variables for the first-order scalars and vector in the metric in the Poisson gauge, 
\begin{subequations}
\begin{align}
\phi_1 &\equiv -\left(U_{1\rm N}+\frac{2}{c^2}U_{1\rm P}\right) \,, \\  
\psi_1 &\equiv -\left(V_{1\rm N}+\frac{2}{c^2}V_{1\rm P}\right) \,,\\ 
 \omega_{1i} &\equiv B_{1i}^{\rm N} +\frac{1}{c^3}B_{1i}^{\rm P} \,,
\end{align}
and introduce the resummed expressions for the gauge generators,
\begin{align}
  \alpha_1  &= \xi^0_1 + \frac{1}{c^2} \chi^0_1 \,, \\
  \beta_1   &= {\cal S}_1 + \frac{1}{c^2} \Sigma_1 \,, \\ 
   d_{1i} &= {\cal S}_{1i_\perp} + \frac 1 {c^2} \Sigma_{1i_\perp} \,, 
\end{align}
\end{subequations}
where we have decomposed the spatial displacement in scalar and vector contributions as ${\cal S}_{1i} = {\cal S}_{1,i} + {\cal S}_{1i_\perp}$ and $ \Sigma_{1i}  = \Sigma_{1,i} + \Sigma_{1i_\perp}$. Note that in virtue of Eq.\,(\ref{vpoisLrep}) we have the following correspondence
\begin{align}
\frac{\dot{\beta}_1}{c}&=\frac{v_1}{c} \\
\frac{\dot{d}_{1i}}{c}&=\frac{\dot{{\cal S}}_{1i_\perp}}{c} + \frac 1 {c^3} \dot{\Sigma}_{1i_\perp}=\frac{v_{1i_\perp}}{c} \,.  \label{mapvel1}
\end{align}
Let us now rewrite Eqs.\,\eqref{u0trafo1}, \eqref{uitrafo1},  
and~\eqref{gijtrafo1} in terms of these resummed variables. They become respectively
\begin{subequations}
\begin{align}
0&=-\frac{\phi_1}{c^2}-\frac{ {\cal H} \alpha_1}{c^2} -\frac{ \dot \alpha_1}{c^2} \,, \label{PF1a}\\
\alpha_{1,i} &=   \dot \beta_{1,i} \,, \label{PF1b}\\
 \frac{C_{1i}^\perp}{c\,a} &=-\frac{ \omega_{1i}}{c^3} +\frac{\dot d_{1i}}{c} \,,  \label{PF1Cperp}\\
g_{1ij_{\rm L}}&=a^2\Bigg \{\left[1-2\left(\frac{\psi_1}{c^2} - \frac{{\cal H} \alpha_1}{c^2}\right)\right]\delta_{ij}  \nonumber\\ 
 &\quad +2\beta_{1,ij}+2d_{1(i,j)} \Bigg\} \,. \label{PFgij1} 
\end{align}
\end{subequations}
These equations are identical with the fully GR gauge transformation at first order, see Eqs.\,\eqref{app:psi1}, \eqref{B1}, \eqref{omega1}, and~\eqref{gijL1}, given in Appendix~\ref{app:linear PT}.
Thus, we have shown that we recover first-order relativistic perturbation theory from the PF approach.
Evidently, to achieve this matching between these different perturbation approaches, 
the introduction of the above resummed variables is essential.

We now solve the above equations by making use of the first-order results in the Poisson gauge that we reported in Sec.\ \ref{sec:EulerLinear}. From Eqs.\,\eqref{PF1a} and~\eqref{PF1b}, the solutions for the time-gauge generator and the scalar in the spatial transformation read, \cite{Villa:2015ppa}
\begin{eqnarray}
\alpha_{1} &=& 
    -\frac{2}{3} \frac{\dot{\cal D}\varphi_0}{\stuff} \,, \label{ouralpha1s}\\
\beta_{1} &=& 
   -\frac{2}{3} \frac{{\cal D}\varphi_0}{\stuff} \,. \label{ourbeta1s} 
\end{eqnarray}
By using the first-order solutions in the Poisson gauge [see Eq.\,\eqref{omega1P}], Eq.\,\eqref{PF1Cperp} reads
\begin{equation}
\frac{\dot{d}_{1i}}{c}= - \frac{6 \stuff}{a^2 c^3} \Delta^{-1} C_{1i}^\perp + \frac{C_{1i}^\perp}{ac} \,,
\end{equation}
which coincides of course with Eq.\,\eqref{mapvel1}. The constant $C_{1i}^\perp$ represents the initial vorticity, i.e., $\fett{\nabla} \times \mathbf{C}_1^\perp= 2\,a_{\rm ini}\fett{\omega}^{\rm ini}_1$.  
This can be easily seen by the comparison between Eq.\,\eqref{mapvel1}
\begin{align}
\frac{\dot{d}_{1i}}{c}&=\frac{\dot{{\cal S}}_{1i_\perp}}{c} + \frac 1 {c^3} \dot{\Sigma}_{1i_\perp}  \nonumber\\
&= \frac{C_{1i}^\perp}{ac}- \frac{6 \stuff}{a^2 c^3} \Delta^{-1} C_{1i}^\perp \,,
\end{align}
and the linearization of Eq.\,\eqref{newtvort}
\begin{equation}
 \varepsilon^{ijk} \frac{\dot{{\cal S}}_{1\perp \,k,j}}{c}  = \frac{a_{\rm ini}}{ac} \omega_{1{\rm ini}}^i \,.
\end{equation}
Finally $d_{1i}$ is given by time integration of
\begin{equation}
\frac 1 2 \frac{\dot{d}_{1i}}{c}= - \frac{6 \stuff}{a^2 c^3} \Delta^{-1} C_{1i}^\perp + \frac{C_{1i}^\perp}{ac} \,.
\end{equation}

Concluding, we obtain for the shift of the metric 
\begin{equation}
g_{0i_{1\rm L}}= 2 a \,a_{\rm ini} \frac{\omega_{1i}^{\rm ini}}{c} \,,
\end{equation}
and for the spatial metric we find
\begin{align}
g_{ij_{1\rm L}}&=a^2\left[ \left(1-2 \frac{g\,\varphi_0}{c^2}\right)\delta_{ij}
 + \dfrac{4}{3} \dfrac{{\cal D}\varphi_{0\,,ij}}{\mathcal{H}_0^2\Omega_{\rm m0}}+2 d_{1(i,j)} \right] \,.
\end{align}
These results are, to our knowledge, new. 
Let us finally make a comment about the linear frame dragging resulting in the Lagrangian gauge. At first order, the gauge invariant definition of the frame-dragging potential is \cite{Bardeen:1980kt} 
\begin{equation}
\Psi_{1i}= \omega_{1i}-\dot{F}_{1i}\,.
\end{equation}
In the Poisson gauge the frame dragging is given by the shift $\omega_{1i}$ [see Eq.\,\eqref{omega1P}] whereas in the Lagrangian gauge it is given by the above combination between the shift and the time derivative of the vector mode in the spatial metric.
The result for the gauge invariant potential is of course the same and reads $\Psi_{1i}=- 6 \stuff \Delta^{-1}C_{1i}^\perp /(a^2 c^2)$, where $\nabla \times \fett{C}_1^{\perp}= 2\,a_{\rm ini} \fett{\omega}^{\rm ini}_1$.


\subsection{The 1PF Lagrangian metric}

According to paper I \cite{Milillo:2015cva}, the 1PF variables are the resummed variables including the first relativistic corrections, in the $1/c$ expansion, to Newtonian variables. When Einstein equations in the Poisson gauge are written in terms of these variables, i.e., up to the 1PF order, they reproduce both the Newtonian equations in the Eulerian approach and relativistic PT equations at linear order in the Poisson gauge. In other words, in paper I \cite{Milillo:2015cva} the 1PF  variables are constructed by a careful analysis of the Einstein equations in the Poisson gauge. By contrast, in the present paper we do not derive the Einstein equations in the Lagrangian gauge. We obtain the results for the Lagrangian metric via a gauge transformation from the Poisson gauge. Nevertheless we showed that we are able to recover both the Newtonian limit and relativistic PT at linear order. Therefore, since we have already established what we need for the Newtonian limit and for first order in PT, we are able to write down the corresponding metric by using the space-space components of the transformation rule for the metrix tensor. We find
\begin{widetext}
\begin{align}\label{gijtrafo0PF}
g_{ij_{\rm L}} &= a^2\bar{\cal J}^k\,\!_i \bar{\cal J}^n\,\!_j \delta_{kn} +\frac{a^2}{c^2}\left[ 2\Sigma^k_{,(i} \bar{\cal J}_{kj)} -\xi^0_{,i} \xi^0_{,j} +\left(2V_{\rm N}+2\mathcal{H}\xi^0\right)\bar{\cal J}^k\,\!\!_i\bar{\cal J}^n\,\!\!_j \delta_{kn} \right]   \nonumber \\ 
& +\frac{a^2}{c^4}\Bigg\{  \xi^0_{,i} \xi^0_{,j} \left(2U_{\rm N}-2\mathcal{H}\xi^0\right)- 2 \xi^0_{,i}\bar{\cal J}^k\,\!_j B^{\rm N}_k -  \xi^0_{,(i}\chi^0_{,j)} 
 +4  \Sigma^k_{,(i} \bar{\cal J}^n\,\!_{j)} \left( V_{\rm N} + \mathcal{H}\xi^0\right)\delta_{kn}
\nonumber \\
&  + \bar{\cal J}^k\,\!_i \bar{\cal J}^n\,\!_j\left[\left(2V_{\rm N}^2+4V_{\rm P}+4\mathcal{H}\xi^0 V_{\rm N}+ 2\xi^0 \dot{ V}_{\rm N}+2 \frac{\partial V_{\rm N}}{\partial x^i}\Bigg|_{x^i=q^i+{\cal S}^i} \Sigma^i+\mathcal{H}\xi^{0^2}+\mathcal{H}\chi^0\right)\delta_{kn}+h_{kn}\right]  \Bigg\} \,.
\end{align}
This is our final result.
\end{widetext}


\section{Conclusions}\label{sec:concl}

In this paper we consider the PF approximation scheme, recently introduced in paper I \cite{Milillo:2015cva} to study cosmic structure formation on all scales from the perspective of GR. 
The PF approach provides nonlinear GR corrections to Newtonian dynamics on small scales and, when linearized, it recovers relativistic perturbation theory, which is the leading-order description on large scales.
In paper I \cite{Milillo:2015cva} the PF formalism was developed in the Poisson gauge where the Einstein equations at the leading order in the PF approximation reduce to the Newtonian equations in the Eulerian formulation of cosmic fluid dynamics. In the present work we develop the PF approach in the Lagrangian-coordinates formulation.

We consider a vortical and pressureless fluid in a flat $\Lambda$CDM universe.
In the context of Newtonian theory, we first review the fully nonlinear equations in the Eulerian approach, and then derive the corresponding equations in the Lagrangian approach  (section \ref{sec:Newton}).
In the Lagrangian approach, the displacement field is the only dynamical quantity, and for its three components we derive novel evolution equations. Specifically, for the scalar part of the displacement we derive the generalized Bernoulli equation, which could be viewed as the statement of energy conservation of a vortical fluid.
For the two vector components of the displacement, we derive the so-called Cauchy invariants, which state the invariance of the fluid Lagrangian under relabeling symmetry (for an extensive discussion see Ref.\,\cite{FrischVillone}). 
These Newtonian equations are new and were previously known only for some noncosmological fluids (see, e.g., \cite{Zheligovsky:2013eca}).

The remaining part of the paper deals with GR corrections to the Newtonian results within the PF approach in Lagrangian coordinates.
From the relativistic point of view, the description of a vortical fluid is complicated by the fact that it is not possible to use the most natural (and most popular) gauge related to the Lagrangian approach, namely the SCO gauge. The latter, in fact, can only be defined if the matter is pressureless (as in our case) and also irrotational.
We therefore construct a new gauge which is suitable for the Lagrangian description of the dynamics of vortical dust, the Lagrangian gauge (Sec.\ \ref{sec:LagrangeGauge}).  Note that in the limit of vanishing vorticity, this Lagrangian gauge becomes identical with the SCO gauge. Furthermore, let us remark that we derive the definition of this Lagrangian gauge from a fully nonperturbative perspective, by exploiting the dust approximation, i.e., the fact that the four-velocity of the dust satisfies the geodesic equation and the exact expression for vorticity in GR.

To obtain the Lagrangian-coordinates approach in the PF scheme, we perform a gauge transformation from Poisson gauge to Lagrangian gauge (Sec.\ \ref{ssec:L3}). We choose to proceed in this way to highlight the physical interpretation of this  gauge transformation: fairly similar to the spatial transformation from Eulerian to Lagrangian coordinates in Newton theory, the outlined gauge transformation amounts to a four-dimensional coordinate transformation involving, to the leading order in 1/c, the fully nonlinear Newtonian displacement field in the spatial component of the gauge transformation, and the fully nonlinear Newtonian velocity potential in the temporal part of the gauge transformation. We find that, to the leading order in the PF scheme, the corresponding constraints from the gauge transformation yield the generalized Bernoulli equation and the Cauchy invariants, thus establishing the Newtonian results in the leading order and proving the consistency of our introduced approximation scheme.

Our formalism could be used to implement GR corrections in $N$-body simulations. Such simulations are traditionally the standard tool to study the process of cosmic structure formation, which however requires the validity of the Newtonian approximation. 
Implementing GR corrections in such codes would be important in order to achieve the target of 1\% accuracy of such simulations (required for future galaxy surveys)  \cite{Schneider:2015yka}, especially considering that on scales of the order of the Hubble horizon, causality, retardation and other GR effects may become important.
First steps in this direction have been recently made in Refs.\,\cite{Bruni:2013mua,Thomas:2015kua} where the authors extracted the frame-dragging gravitomagnetic vector potentials from Newtonian simulations. Their analysis however relied on a Eulerian description, so one straightforward application of our formalism would be to repeat their analysis but calculate the frame dragging within the Lagrangian approach.
 Furthermore, initial conditions for Newtonian simulations are usually set up using the Zel'dovich approximation, or its second-order extension 2LPT (see \cite{Bernardeau:2001qr} and references therein). Both of these approximation schemes are Lagrangian too, however, only Newtonian, so one straightforward application would be to generalize 2LPT within the Lagrangian PF approach. Similar considerations to this have been recently made in \cite{Rampf:2014mga,Fidler:2015npa,Christopherson:2015ank,Fidler:2016tir} where however relativistic perturbation theory has been used instead of the PF scheme.

Finally, we note that recently the first GR  cosmological numerical simulations have been produced. In these simulations the space-time metric is self-consistently calculated by integrating Einstein equations, either with an $N$-body approach within the weak-field approximation \cite{Adamek:2013wja,Adamek:2015eda}, or assuming an irrotational pressureless fluid in fully nonlinear numerical relativity \cite{Giblin:2015vwq,Bentivegna:2015flc,Mertens:2015ttp}. In particular, in  Refs.\  \cite{Giblin:2015vwq,Bentivegna:2015flc,Mertens:2015ttp} the SCO gauge has been used, and therefore the Lagrangian PF approximation introduced here would be an ideal approximate framework to compare with these fully nonlinear GR results, for instance to help  establish the relevance of the relativistic corrections to the Newtonian results.

\section*{ACKNOWLEDGMENTS}

We thank Thomas Buchert,  Uriel Frisch, Toshifumi Futamase and Barbara Villone for useful discussions.
C.R.\ acknowledges the support of the individual Grants No.\ \mbox{RA 2523/1-1} and No.\ RA 2523/1-2 from the Deutsche Forschungsgemeinschaft (DFG). 
E.V.\ thanks ``Fondazione Angelo della Riccia," the University of Portsmouth and the INFN-INDARK initiative Grant No.\ IS PD51 for financial support.  
During the preparation of this work D.B.\ was supported by the DFG through the Transregio 33, The Dark Universe.
M. B. was supported by the UK STFC Grants No.\ ST/L005573/1, No.\ ST/K00090X/1 and No.\ ST/ N000668/1.


\appendix

\section{First-order gauge transformation}\label{app:linear PT}

In this appendix we provide some first-order calculations. For notational simplicity, we use
 a comma to denote a spatial partial derivative ($\partial_i =_{,i}$), and a dot denotes the partial derivative with respect to conformal time.
The components of a spatially flat FLRW metric perturbed up to first order are written in any gauge as 
\begin{subequations}
   \label{metricIIany}
\begin{align} 
    g_{00}&=-a^2\left( 1+2 \phi_1 \right) \\
    g_{0i}&=a^2 \left(B_{1,i}-\omega_{1i} \right) \\
    g_{ij}&=a^2\left[ \left(1-2\psi_{1}\right)\delta_{ij}+2D_{ij}E_{1}+2F_{1(i,j)}\right] \,,
\end{align}
\end{subequations}
where  we make use of the operator $D_{ij}\equiv\partial_i\partial_j-(1/3)\nab^2\delta_{ij}$,  $\omega_{1i}$ and $F_{1i}$ are transverse vectors, i.e., 
$\delta^{ij}\omega_{1i,j}=\delta^{ij}F_{1i,j}=0$, and we neglect first-order tensor modes in the spatial metric.
The background part (which is by definition only time dependent) is given by
\begin{equation} \label{scalef}
\overline{g}_{00}=-a^2(\eta) \,, \qquad\qquad \overline{g}_{ij}=a^2(\eta)\,\delta_{ij} \,.
\end{equation}
As before, $\eta$ is the conformal time ($a\, \dd \eta = \dd t$, where $t$ is the cosmic time), and $a(\eta)$ the FLRW scale factor, which obeys the Friedmann equations.

The four-velocity of matter is $u^\mu=  {\rm d}x^\mu / (c {\rm d}\tau)$, where $\tau$ is the proper (comoving) time, comoving with the fluid. To first order we have
\begin{equation} \label{umu}
u^\mu=\frac{1}{ac}\left(\delta^\mu_0+u^\mu_1\right),
\end{equation}
where $u_1^\mu$ is the first-order peculiar velocity (peculiar in the spatial and temporal sense).
From the normalization condition $u^\mu u^\nu g_{\mu\nu}=-1$, we 
obtain the constraint for the time component of $u^\mu$, which reads up to first order (in any gauge)
\begin{equation} \label{v0any}
u^0_1 =  -\frac{\phi_1}{c^2}\,.
\end{equation}
The perturbations of the spatial components $v^i$ split as usual in scalar and vector parts
\begin{equation} \label{velocity}
 u^i _{1}= \delta^{ij}v_{1,j}+v^i_{1\perp}\,,
\end{equation}
where $v^i_{1\perp,i}=0$.
Finally, the perturbation in the matter density up to first order is written as $\rho=\overline{\rho}+\rho_1$, where the background density $\overline{\rho}$ is time dependent only. The density contrast is defined by
\be \label{def:delta}
  \delta_1(\eta, \fett{x}) \equiv \frac{\rho_1(\eta,\fett{x})  -\bar \rho(\eta) }{\bar \rho (\eta)}  \,.
\ee

\subsection{First-order gauge transformations} 
We specialize the first-order transformation rules to the specific gauge transformation considered in this paper, namely the transformation from the Poisson gauge, with coordinates $x^\mu(q^\nu)$, to the Lagrangian gauge, with coordinates $q^\mu(q^\nu)$. As before, quantities in the Lagrangian gauge are indicated with a subscript ``L''.
Following Ref.~\cite{Villa:2014aja}, we adopt the so-called passive approach, where the gauge transformation is seen as a coordinate transformation $q^\mu \rightarrow x^\mu(q^\nu)$, where $q^\mu$ are coordinates in the Lagrangian gauge, and $x^\mu$ the coordinates in the Poisson gauge. 
Up to first order the temporal and spatial gauge transformations are
\begin{align} \label{TdG}
  \tau &= \eta - \frac{\xi^0_{1}}{c^2}  \,,\\
  q^i &= x^i-\xi^i_{1}\,,
\intertext{with inverse} 
 \label{TdGi}
  \eta &= \tau + \frac{\xi^0_{1}}{c^2}   \,,\\
  x^i &= q^i+\xi^i_{1}\,,
\end{align}
where all the quantities are evaluated at the same point on the background space-time where the coordinates $x^\mu$ and $q^\mu$ coincide. As usual, the four vectors $\xi_{1}^\mu$ can be decomposed into scalar and vector parts
\begin{equation} \label{xidecomp}
\xi^0_{1} = \alpha_{1}\, , \qquad \xi^{i}_{1} =\delta^{ij} \beta_{1,j} + d^i_{1}\, , \qquad  \text{with} \quad  d^i_{1,i} = 0\,.
\end{equation}

In the Poisson gauge the space-time metric perturbed up to first order is given by
\begin{align} 
    g_{00}&=-a^2\left( 1+2 \frac{\phi_1}{c^2} \right) \,, \\
    g_{0i}&=-a^2\frac{\omega_{1i}}{c^3} \,, \\
    g_{ij}&=a^2\left[ \left(1-2\frac{\psi_{1}}{c^2}\right)\delta_{ij}\right] \,,  
\end{align}
and in the Lagrangian gauge we have
\begin{align} 
    g_{00}&=-a^2\,, \\
    g_{0i}&=-a^2\frac{\omega_{1i_{\rm L}}}{c} \,, \\
    g_{ij}&=a^2\left[ \left(1-2\psi_{1_{\rm L}}\right)\delta_{ij}+2D_{ij}E_{1_{\rm L}}+2F_{1_{\rm L}\,(i,j)}\right]\,, 
\end{align}
where $\omega_{1i_{\rm L}}=-C_{i}^\perp$ from our gauge condition \eqref{gaugecondshift}.

\paragraph{Metric tensor.}
We find the following first-order transformations for the metric tensor
\begin{itemize}
\item scalar perturbations
\begin{eqnarray}
 0&=& \frac{\phi_1}{c^2} + \frac{{\cal H} \alpha_1}{c^2} +\frac{ \dot \alpha_1}{c^2} \,,  \label{app:psi1}\\ 
\psi_{1_{\rm L}} &=& \frac{\psi_1}{c^2} - \frac{{\cal H} \alpha_1}{c^2} -\frac{1}{3}\nab^2\beta_1 \,,  \label{phi1}\\ 
 \alpha_1 &=&   \dot \beta_1 \,, \label{B1}\\
 E_{1_{\rm L}} &=& \beta_1 \,, \label{E1}
\end{eqnarray}
\item vector perturbations 
\begin{eqnarray}
\frac{ \omega_{1i_{\rm L}}}{c}  &=&\frac{ \omega_{1i}}{c^3} -\frac{\dot d_{1i}}{c} \,, \label{omega1}\\
 F_{1i_{\rm{L}}} &=&  d_{i_1} \,, \label{F1} 
\end{eqnarray}
\end{itemize}
where the dot denotes partial derivative with respect to conformal time and ${\mathcal H}=\dot{a}/a= a H$ is the conformal Hubble parameter. 

Putting all the results together, the spatial metric in the Lagrangian gauge is given by
\begin{align}\label{gijL1}
g_{ij_{\rm L}}&=a^2\! \left[\left\{ 1-2\left(\frac{\psi_{1_{\rm L}}}{c^2} - \frac{{\cal H} \alpha_1}{c^2} \right) \! \right\} \delta_{ij} 
  +2\beta_{1,ij}+ 2d_{1(i,j)} \right] .
\end{align}

\paragraph{Three-velocity.} 
The transformation of the temporal part of the peculiar velocity is obtained from Eq.\,\eqref{v0any} and reads
\begin{equation}\label{v01}
0=-\frac{\phi_1}{c^2}-\frac{ {\cal H} \alpha_1}{c^2} -\frac{ \dot \alpha_1}{c^2}\,.
\end{equation}
For the scalar and vector part of the spatial peculiar velocity we find
\begin{equation}\label{v1}
0=\frac{v_1}{c}-\frac{\dot \beta_1}{c}  \,,
\end{equation}
and
\begin{equation}\label{w1}
0=\frac{v_{i_{1\perp}}}{c}-\frac{\dot d_{i_1}}{c}\,. 
\end{equation}
\paragraph{Matter density.}
Finally, the perturbation of the density contrast transforms as
\begin{equation}\label{rho1}
\delta_{1_{\rm L}}=\delta_1 - \frac{3 {\cal H}\,\alpha_1}{c^2}\,.
\end{equation}


\end{document}